\documentclass[apj]{emulateapj}
\usepackage{natbib}
\defcitealias{tremonti04a}{T04}
\defcitealias{cl01a}{CL01}
\newcommand{\kms}{\ifmmode \mathrm{km~s^{-1}}\else km~s$^{-1}$\fi}
\newcommand{\smpy}{\ifmmode M_\sun~\mathrm{yr}^{-1}\else M$_\sun$~yr$^{-1}$\fi}
\newcommand{\lir}{\ifmmode L_\mathrm{IR}\else $L_\mathrm{IR}$\fi}
\newcommand{\lsun}{\ifmmode L_\sun\else $L_\sun$\fi}
\newcommand{\msun}{\ifmmode M_\sun\else $M_\sun$\fi}
\newcommand{\zsun}{\ifmmode Z_\sun\else $Z_\sun$\fi}
\newcommand{\htwo}{\ion{H}{2}}

\newcommand{\nt}{\ifmmode $[\ion{N}{2}]$\else[\ion{N}{2}]\fi}
\newcommand{\ntl}{\ifmmode $[\ion{N}{2}]$ \lambda6583\else[\ion{N}{2}]$\lambda6583$\fi}

\newcommand{\oo}{\ifmmode $[\ion{O}{1}]$\else[\ion{O}{1}]\fi}
\newcommand{\ool}{\ifmmode $[\ion{O}{1}]$ \lambda6300\else[\ion{O}{1}]$\lambda6300$\fi}
\newcommand{\otw}{\ifmmode $[\ion{O}{2}]$\else [\ion{O}{2}]\fi}
\newcommand{\otwl}{\ifmmode $[\ion{O}{2}]$ \lambda\lambda3726,~3729\else[\ion{O}{2}]$\lambda\lambda3727,~3729$\fi}
\newcommand{\ot}{\ifmmode $[\ion{O}{3}]$\else[\ion{O}{3}]\fi}
\newcommand{\otl}{\ifmmode $[\ion{O}{3}]$ \lambda5007\else[\ion{O}{3}]$\lambda5007$\fi}
\newcommand{\otll}{\ifmmode $[\ion{O}{3}]$ \lambda\lambda4959,~5007\else[\ion{O}{3}]$\lambda\lambda4959,~5007$\fi}
\newcommand{\st}{\ifmmode $[\ion{S}{2}]$\else[\ion{S}{2}]\fi}
\newcommand{\stl}{\ifmmode $[\ion{S}{2}]$ \lambda\lambda6716,~6731\else[\ion{S}{2}]$\lambda\lambda6717,~6731$\fi}
\newcommand{\rtt}{\ifmmode \mathrm{R}_{23}\else R$_{23}$\fi}
\newcommand{\ott}{\ifmmode \mathrm{O}_{32}\else O$_{32}$\fi}
\newcommand{\te}{\ifmmode T_e\else $T_e$\fi}
\newcommand{\mz}{\ifmmode M-Z\else $M-Z$\fi}
\newcommand{\lz}{\ifmmode L-Z\else $L-Z$\fi}
\newcommand{\kp}{\ifmmode K^\prime\else $K^\prime$\fi}
\newcommand{\hkp}{\ifmmode H-K^\prime\else $H-K^\prime$\fi}

\shorttitle{Abundances of Luminous Infrared Galaxies}
\shortauthors{Rupke, Veilleux, \& Baker}

\begin{document}

\journalinfo{ApJ, 1 Dec 2007, v670n2, in press}

\title{The Oxygen Abundances of Luminous and Ultraluminous Infrared
  Galaxies}

\author{David S. N. Rupke, Sylvain Veilleux} \affil{Department of
  Astronomy, University of Maryland, College Park, MD 20742-2421}
\email{drupke@astro.umd.edu}

\author{and Andrew J. Baker} \affil{Department of Physics and
  Astronomy, Rutgers, the State University of New Jersey, 136
  Frelinghuysen Road, Piscataway, NJ 08854-8019}

\begin{abstract}

  Luminous and ultraluminous infrared galaxies (LIRGs and ULIRGs)
  dominate the star formation rate budget of the universe at $z \ga
  1$, yet no local measurements of their heavy element abundances
  exist.  We measure nuclear or near-nuclear oxygen abundances in a
  sample of 100 star-forming LIRGs and ULIRGs using new, previously
  published, and archival spectroscopy of strong emission lines
  (including \otwl) in galaxies with redshifts $\langle z \rangle \sim
  0.1$.  When compared to local emission-line galaxies of similar
  luminosity and mass (using the near-infrared luminosity-metallicity
  and mass-metallicity relations), we find that LIRGs and ULIRGs are
  under-abundant by a factor of two on average.  As a corollary, LIRGs
  and ULIRGs also have smaller effective yields.  We conclude that the
  observed under-abundance results from the combination of a decrease
  of abundance with increasing radius in the progenitor galaxies and
  strong, interaction- or merger-induced gas inflow into the galaxy
  nucleus.  This conclusion demonstrates that local abundance scaling
  relations are not universal, a fact that must be accounted for when
  interpreting abundances earlier in the universe's history when
  merger-induced star formation was the dominant mode.  We use our
  local sample to compare to high-redshift samples and assess
  abundance evolution in LIRGs and ULIRGs.  We find that abundances in
  these systems increased by $\sim$0.2~dex from $z\sim0.6$ to
  $z\sim0.1$.  Evolution from $z\sim2$ submillimeter galaxies to
  $z\sim0.1$ ULIRGs also appears to be present, though uncertainty due
  to spectroscopic limitations is large.

\end{abstract}

\keywords{galaxies: abundances --- galaxies: evolution --- galaxies:
  interactions --- galaxies: ISM --- galaxies: kinematics and dynamics
  --- infrared: galaxies}


\section{INTRODUCTION} \label{sec:introduction}

Mid-infrared and submillimeter observations show that luminous and
ultraluminous infrared galaxies (LIRGs and ULIRGs)\footnote{LIRGs are
  defined by $10^{11} < \lir/\lsun < 10^{12}$ and ULIRGs by $10^{12} <
  \lir/\lsun < 10^{13}$, where \lir\ is the `total' infrared
  luminosity from $8-1000$~\micron.} host most of the star formation
in the universe at $z \ga 1$
\citep{lefloch05a,chapman05a,daddi05a,wcb06a,caputi07a}.
Understanding local examples of these sources is thus a window to star
formation and accretion onto supermassive black holes at the epoch of
highest star formation rate and active galactic nucleus (AGN) density
\citep[e.g.,][]{mpd98a,ssg95a}.

A great deal is known about nearby ULIRGs; for a recent review see
\citet{lfs06a}.  All ULIRGs possess strong starbursts, and many also
host optically-visible AGN \citep{vks99a}.  The starbursts and AGN are
on average heavily obscured \citep{genzel98a,hao07a}, and
optically-invisible AGN may be present \citep{lvg99a,armus07a}.  The
prevalence of optically-visible AGN increases with increasing infrared
luminosity \citep{vks99a} and as the merger progresses \citep{vks02a}.
It is thus hypothesized that ULIRGs play a role in the evolution of
quasars.  When obscuring dust is removed from a buried AGN (by
starburst- or AGN-driven outflows; see, e.g.,
\citealt{rvs05c,rvs05b}), a bright quasar is left \citep{sanders88a}.
This hypothesis is under scrutiny; comparison to the fundamental plane
of ellipticals shows that many ULIRGs are evolving into moderate-mass
ellipticals \citep{genzel01a,vks02a,tacconi02a,dasyra06a,dasyra06b},
similar to but slightly less massive than the hosts of optically
bright quasars \citep{dasyra07a}.

LIRGs host less intense starbursts than ULIRGs, and the frequency of
occurrence of optically-visible AGN in LIRGs is much smaller
\citep{veilleux95a}.  Photometric studies divide LIRGs into two groups
\citep{ishida04a}.  The most luminous are early in the merger sequence
of two roughly equal-mass galaxies, and thus may be the progenitors of
ULIRGs \citep{arribas04a,ishida04a}.  Other LIRGs are unequal-mass
mergers or isolated disk galaxies which may or may not be experiencing
an interaction \citep{ishida04a}.  Kinematic studies confirm that the
pair mass ratios in at least some LIRG interactions are of order $1-3$
\citep{rj06b,dasyra06a}.

The interstellar medium in LIRGs and ULIRGs is in a kinematically
extreme state, dominated by inflows \citep[e.g.,][]{bh96a,mh96a},
outflows
(\citealt{heckman00a,rvs02a,rvs05c,rvs05a,rvs05b,martin05a,martin06a};
see \citealt{vcb05a} for a recent review), and turbulent motions
\citep{ds98a}.  These gas motions have the power to significantly
alter the chemical states of the progenitor galaxies
\citep[e.g.,][]{edmunds90a,ke99a,dalcanton07a}.  Ongoing, intense star
formation in LIRGs and ULIRGs is also producing and redistributing
heavy metals at a prodigious rate.  In this paper we describe the
first comprehensive study of the oxygen abundances of local LIRGs and
ULIRGs (\S\S $\ref{sec:sample-selection}-\ref{sec:abundances}$).  This
will allow us to assess the effects of these processes on the
gas-phase abundances of infrared-selected, interacting galaxies.
Studies of optically-selected mergers suggest that gas motions do
alter nuclear abundances \citep{kewley06b}.  Here we present evidence
that this is true also of strong mergers with high star formation
rates.

In order to understand the chemical evolution of LIRGs and ULIRGs, it
is crucial to compare these sources to weakly- or
non-interacting galaxies with modest star formation, which represent 
the progenitors of LIRGs and ULIRGs.  To this end,
we compare LIRGs and ULIRGs to published luminosity-metallicity,
mass-metallicity, and mass-effective yield relations (\S\S
$\ref{sec:lzreln}-\ref{sec:effective-yield}$).  We discuss these
results in \S\ref{sec:discussion}.

There exist a few measurements of oxygen abundances in high-redshift
infrared-luminous galaxies.  \citet{liang04a} measure the abundances
of $\sim$20 $z=0.4-0.9$ LIRGs selected at 15~\micron\ by the {\it
  Infrared Space Observatory} ({\it ISO}).  Abundance measurements
also exist for a handful of $z\sim2$ submillimeter-selected galaxies
that have total infrared luminosities greater than or equal to those
of ULIRGs \citep{tecza04a,swinbank04a}.  Finally, we describe in this
paper a handful of new moderate-redshift measurements.

While these high-$z$ measurements are valuable on their own, they are
also sparse and, in many cases, uncertain.  The more robust
measurements of local LIRGs and ULIRGs that we report here facilitate
two comparisons: we can now compare the enrichment histories of
infrared-luminous and infrared-faint galaxies in the local universe
(\S\S~\ref{sec:l-z-relation}, \ref{sec:comp-other-merg}, and
\ref{sec:mzreln}); and we can look for evidence of chemical evolution
among infrared-luminous populations as a function of redshift
(\S\ref{sec:comp-high-redsh}).

We summarize our work and discuss its consequences in
\S\ref{sec:summary-outlook}.

Throughout the paper, we use the notation (O/H) to refer to the ratio
of the number densities of O and H atoms in the ISM.  The variable $Z$
refers instead to the mass fraction of O relative to the total mass of
gas.  These two variables are related by a constant: $Z=16/C$(O/H),
where $C\sim1.4$ is the ratio of total to H gas masses.  Where
appropriate, we use (O/H) and $Z$ interchangeably.  For the solar
oxygen abundance, we use the recent value from \citet{asplund04a}:
$12+\mathrm{log(O/H)_\sun}=8.66$.  For all calculations, we assume the
standard cosmology of $H_0 = 75$~\kms~Mpc$^{-1}$, $\Omega_m = 0.3$,
and $\Omega_{\Lambda} = 0.7$.

\section{SAMPLE SELECTION} \label{sec:sample-selection}

Abundance diagnostics of star-forming galaxies rely mostly on emission
lines.  The emission lines of many LIRGs and ULIRGs, however, may
include contributions from an AGN and/or strong shocks
\citep{kvs98a,vks99a}.  Abundance diagnostics are generally calibrated
using galaxies with modest star formation, whose emission line fluxes
do not contain strong contributions from either of these ionization
mechanisms.  Thus, choosing galaxies whose lines are
starburst-dominated is important for computing accurate abundances.

This decision is complicated by the multiple options for defining a
star-forming galaxy in the phase space of optical emission line flux
ratios.  For instance, using the classic diagnostics of \citet{vo87a},
60\% of LIRGs and one-third of all ULIRGs are `\htwo-region-like'
galaxies \citep{veilleux95a,vks99a}.  More recent work updates this
empirical classification scheme using $\sim$10$^5$ galaxies from the
Sloan Digital Sky Survey (SDSS; \citealt{kauffmann03a,kewley06a}).
Many LIRGs and ULIRGs classified as \htwo\ galaxies using the
\citet{vo87a} scheme lie away from the bulk of local star-forming
galaxies; instead, they lie in the region of the \otl/H$\beta$ vs.
\ntl/H$\alpha$ flux ratio diagram that is between the outer boundary
of the locus of SDSS galaxies \citep{kauffmann03a} and the line
delineating the maximum line ratios achievable by starbursts,
according to theory \citep{kewley01a}.

\begin{figure*}
  \plotone{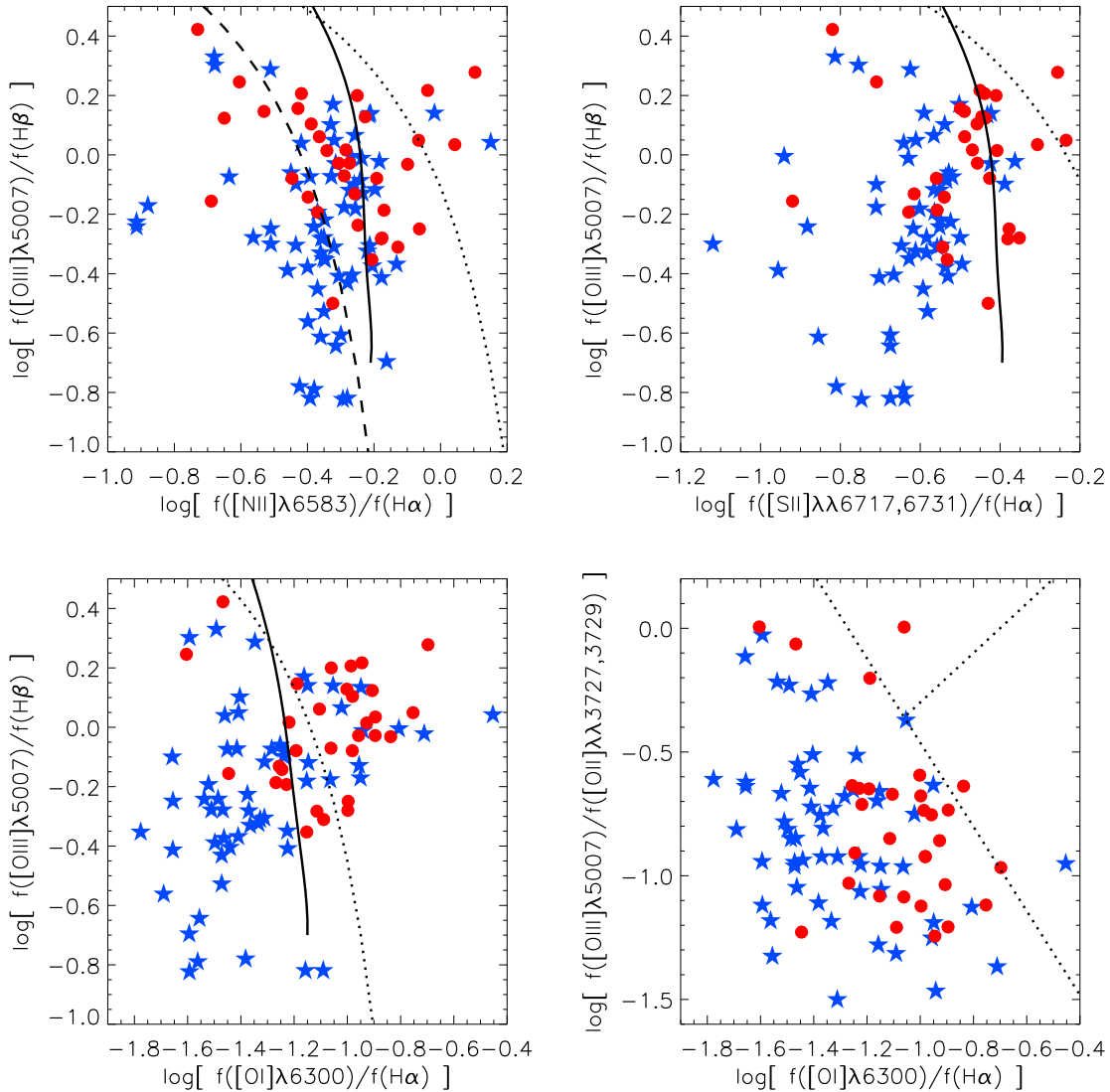}
  \caption{Emission-line ratio diagrams, with fluxes corrected for
    extinction.  Blue stars are luminous infrared galaxies and red
    circles are ultraluminous infrared galaxies.  The solid lines
    separate star-forming galaxies from LINERs \citep{vo87a}; the
    dotted lines denote the phase space limits of the
    \citet{kewley01a} starburst models in all plots except the bottom
    right, where they separate starbursts, LINERs, and AGN
    \citep{kewley06a}; and the dashed line denotes the outer boundary
    of the bulk of star-forming galaxies in the SDSS
    \citep{kauffmann03a}.}
\label{fig_lrats}
\end{figure*}

In \S\ref{sec:uncert-abund-phys}, we discuss in detail the abundance
uncertainties that arise from such physical effects.  For now, we
adopt loose initial selection criteria.  We include in our sample all
galaxies classified as \ion{H}{2} galaxies or low ionization nuclear
emission-line regions (LINERs) under any scheme.  We also require that
line flux uncertainties be smaller than 50\%; almost all fluxes are
far more certain than this, but a few fall near the limit.  In
Figure~\ref{fig_lrats}, we place the galaxies in the current sample on
several line-ratio diagrams, with various classification schemes
over-plotted.

Our ULIRGs are primarily from the 1~Jy sample, which is a complete,
flux-limited, northern-hemisphere sample drawn from the {\it Infrared
  Astronomical Satellite} ({\it IRAS}) database \citep{ks98a}.  We
require that our galaxies have measured \otwl\ fluxes.  The
moderately-high-resolution spectra of \citet{rvs02a,rvs05a}, from Keck
or the MMT, have broad enough wavelength coverage for this purpose.
We supplement this data set with a handful of spectra of 1~Jy objects
from the Fifth Data Release of the SDSS
\citep{york00a,strauss02a,adelman07a}.  To improve our statistics, we
also include a few ULIRGs with published \otw\ fluxes from the Revised
Bright Galaxy Sample (RBGS), the Warm Galaxy Survey (WGS), and the
2~Jy survey \citep{kim95a,wu98a}.  The final local sample of 31 galaxy
nuclei has $\langle z \rangle = 0.14$, with a maximum redshift of
0.27.  However, in addition we include measurements for a few galaxies
with $z = 0.4-0.5$ from the FIRST-FSC catalog
\citep{stanford00a,rvs05a} for the purpose of assessing redshift
evolution (\S\ref{sec:comp-high-redsh}).

We have selected our LIRGs primarily from the Revised Bright Galaxy
Sample, which is also a complete, flux-limited {\it IRAS} sample
\citep{sanders03a}.  Our RBGS data come from a variety of
spectroscopic surveys: (1) \citet{kim95a}; (2) \citet{lk95a}; (3)
\citet{wu98a}; (4) \citet{rvs02a,rvs05b}; (5) the Fourth Data Release
of the SDSS \citep{adelman06a}; and (6) \citet[][nuclear spectra
only]{mk06a}.  Again, to improve our statistics, we include published
fluxes of galaxies from the Warm Galaxy Survey (WGS) and 2~Jy sample
\citep{kim95a,wu98a}.  The sample of 65 galaxy nuclei has $\langle z
\rangle = 0.04$.  (We also have measurements for one LIRG with
$z=0.48$, which we use in \S\ref{sec:comp-high-redsh}.)

The selected galaxies are representative of the local ($z \la 0.2$),
star-forming, infrared-luminous galaxy population.  The 1~Jy sample,
2~Jy sample, RBGS, WGS are complete samples.  The particular galaxies
which appear in this paper are effectively a random selection from
these samples, since they are culled from their parent spectroscopic
studies only on the basis of spectral type and signal-to-noise ratio.
The parent studies from which our spectroscopic data were collated
impose various constraints on sky location, emission-line sensitivity,
and (to a minor degree) redshift.  However, these have no effect on
the physical properties of the galaxies chosen, as shown by
comparisons among Figure~\ref{fig_lrats} and similar figures from the
parent studies \citep[e.g.,][]{kim95a,vks99a}.

Some of the galaxies in our sample are multiple-nucleus
systems\footnote{For some multiple-nucleus systems, only one nucleus
  enters our sample.  This is due to a lack of data or because the
  other nucleus has a Seyfert optical spectral type.}.  For the 1~Jy
ULIRGs, the nuclei are characterized using sensitive optical and
near-infrared imaging \citep{kvs02a,vks02a}.  For other sources, we
specify nuclei using unique designations from the NASA/IPAC
Extragalactic Database (NED; see Table~\ref{tab_samp}).  We treat each
nucleus separately in our analysis, since the apertures for our
spectra are nuclear or near-nuclear (\S\ref{sec:uncert-abund-phys}).
Accordingly, for each multiple-nucleus system we divide the total
system infrared luminosity between component nuclei based on resolved
{\it IRAS} imaging \citep{ssm04a} or the near-infrared luminosities of
the component galaxies (\S\ref{sec:nir-photometry}).  The fraction of
the total infrared luminosity of the system belonging to each nucleus
(or equivalently fractional near-infrared luminosity) for those
systems where the NIR luminosity is used to divide \lir\ is listed in
Table~\ref{tab_samp}.  Seven ULIRG nuclei descend into the LIRG
category due to their nuclear luminosity.  This reclassification does
not impact our results.

In total, there are 100 galaxies or nuclei in our sample.
Table~\ref{tab_samp} lists the basic properties of each galaxy or
nucleus.

The strong emission lines in the \citet{rvs02a,rvs05a} and SDSS
flux-calibrated spectra were measured using the IRAF task SPLOT.  The
average measurement uncertainty in the brightest lines (e.g., \otw\ or
\nt) is 5\% or less.  For weak or noisy lines (\oo, \ot, or \st\ in a
few cases) or those affected by a continuum that has strong stellar
absorption, the uncertainty rises to $\sim20-30$\%.  Table
\ref{tab_lines} lists the line fluxes newly measured for this study.

Our Keck, MMT, and SDSS spectra are typically of high enough
resolution for us to fit simultaneously a Voigt absorption and
Gaussian emission component to H$\beta$.  In the few cases where this
was not possible, we used lower order Balmer lines to estimate the
expected absorption in H$\beta$.  We corrected the H$\alpha$ emission
line in these galaxies for stellar absorption by extrapolating from
the absorption equivalent widths of lower order Balmer lines, using
the relative values for different lines predicted by the oscillator
strengths \citep{menzel69a}:
\begin{equation}
  W_{eq}(\mathrm{H}\alpha)=\frac{W_{eq}(\mathrm{H}_N)}{\sum_{i=3}^{N}f(\mathrm{H}_i)}.
\end{equation}
In this equation, H$_N$ is the Balmer transition from which the
equivalent width of H$\alpha$ is to be calculated; H$_i$ represents
the Balmer transition with upper principal quantum number $i$ (e.g.,
H$_4$ = H$\beta$); and $f(\mathrm{H}_i)$ is its oscillator strength.
\citet{patris03a} tabulate the coefficient values that we used.  Table
\ref{tab_lines} lists the (absorption-uncorrected) H$\alpha$ emission
line and H$\beta$ absorption line equivalent widths for newly measured
data.

The emission lines were corrected for extinction using the Balmer
decrement.  We assume an intrinsic H$\alpha$/H$\beta$ flux ratio of
2.86 \citep{hs87a}, an effective foreground screen, and the starburst
attenuation curve of \citet{calzetti00a}.  Table \ref{tab_lines} lists
the E(\bv) values for newly measured data.

The data quality was checked through cross-correlation of objects
common to different data sets.  The results showed remarkable
consistency, given the number of references from which the data were
drawn.  The typical discrepancy between two measurements of the same
galaxy nucleus is of the order $\sim$0.1~dex.  Where multiple data
were available for a given source, we chose the spectrum with highest
spectral resolution and signal-to-noise.  Exceptions are sources not
drawn from the 1~Jy sample or RBGS that have both published and SDSS
spectra; for consistency, we chose the published data even though the
SDSS data may be of higher spectral resolution or sensitivity.  We
decided to use the SDSS data only to supplement the number of spectra
available from the 1~Jy sample and RBGS (or for data quality checking)
in order to keep the sample selection fairly clean and the size of the
sample manageable.  (I.e., there are a large number of
infrared-luminous galaxies in the SDSS that we did not include; e.g.,
\citealt{pkh05a}.)

\section{ABUNDANCES} \label{sec:abundances}

\subsection{Comparison of Diagnostics} \label{sec:comp-diag}

\begin{figure*}
  \plottwo{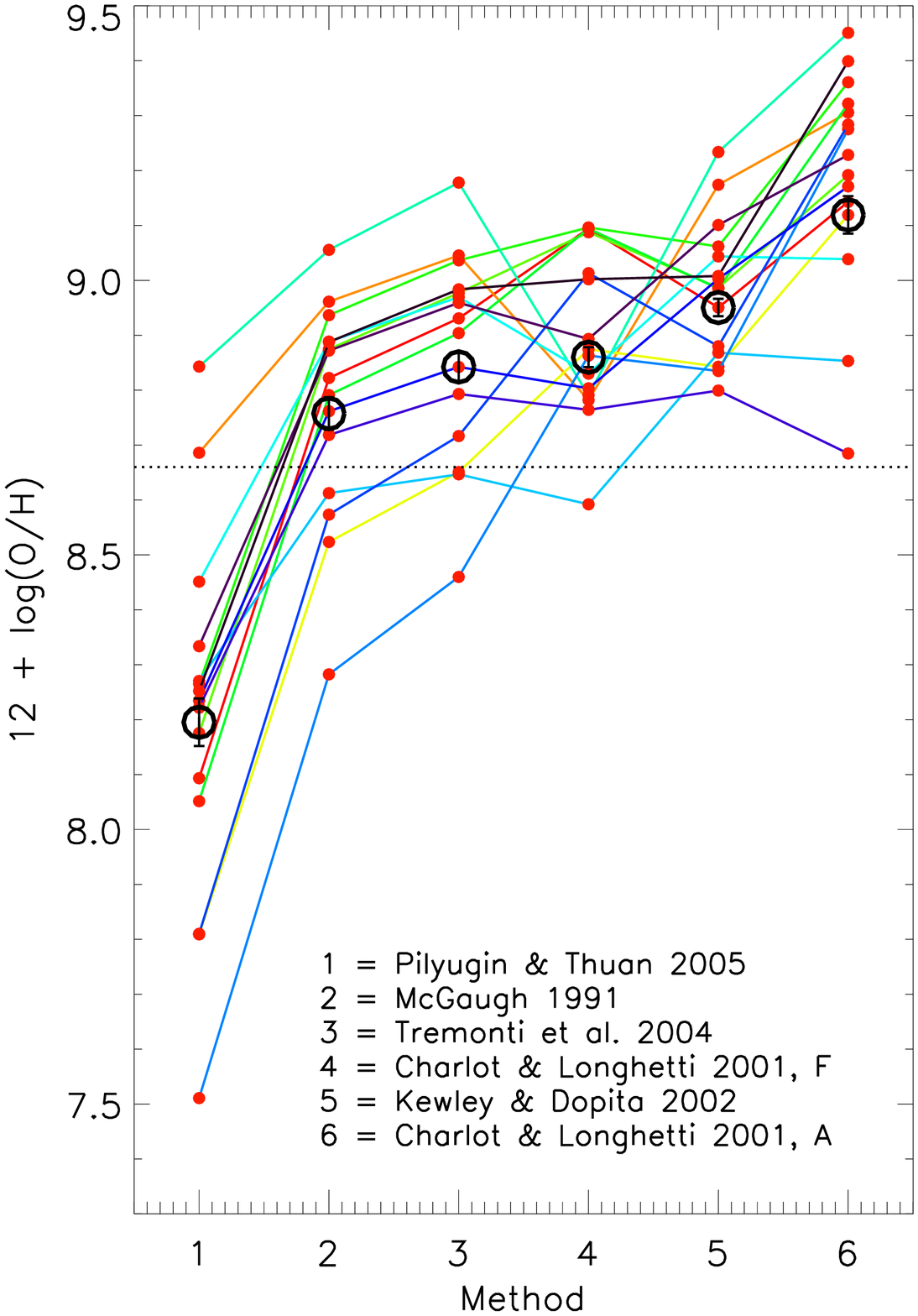}{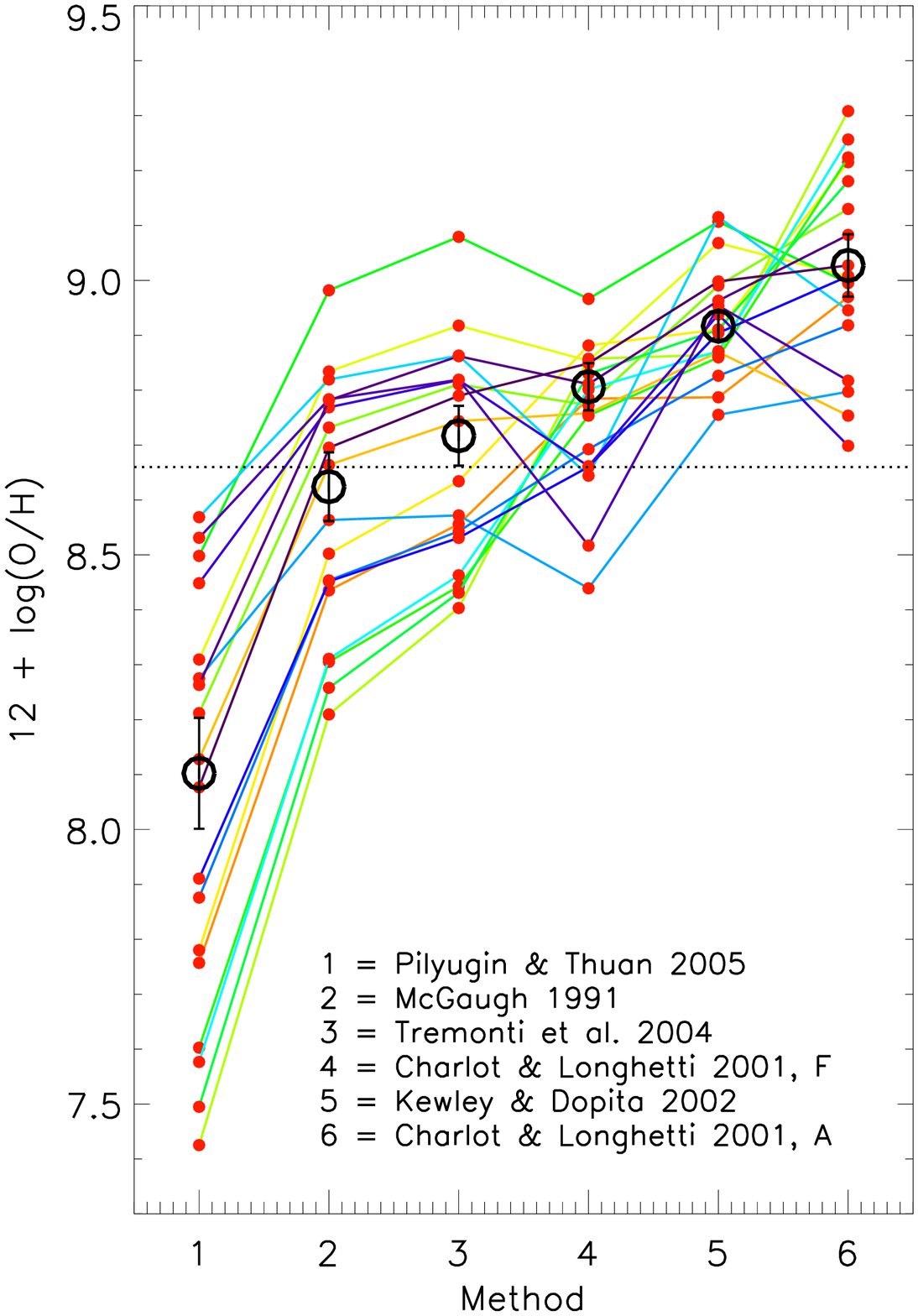}
  \caption{Abundances in six different diagnostic/calibration systems.
    (See \S\ref{sec:comp-diag} for a description of each system.)  We
    select a random sample of LIRGs (left) and ULIRGs (right).
    Colored lines connect different calibrations for the same galaxy.
    The heavy open circles and error bars reflect the median and
    standard error over the entire sample in each system.  We here
    consider only galaxies that pass the second emission-line cut
    (\S\ref{sec:uncert-abund-phys}).}
  \label{fig_zscatt}
\end{figure*}

Numerous strong-line abundance diagnostics exist, each relying on
different combinations of the strongest emission lines measurable in
optical spectra.  Each diagnostic in turn has different absolute
calibrations, based on photoionization models, electron temperature
($T_e$) measurements (i.e., weak-line diagnostics), or a combination
of the two.  Different diagnostics, or different calibrations of the
same diagnostic, can give vastly different abundances for the same
galaxy or group of galaxies.  For instance, models calibrated on measurements
of electron temperatures of \htwo\ regions in nearby galaxies differ
from theoretical calibrations by factors of a few \citep[see,
e.g.,][and references therein]{kbg03a}.

To help the reader understand the uncertainty that attaches to the
choice of a particular diagnostic/calibration combination, we have
investigated several different options.  One of these (the
\citealt{tremonti04a} calibration) we employ in
\S\S~$\ref{sec:lzreln}-\ref{sec:effective-yield}$ to compare our data
with published luminosity-metallicity (\lz), mass-metallicity (\mz),
and mass-effective yield relations.  The others provide a useful
baseline comparison and illustrate some of the uncertainties due to
choice of calibration and physical effects.
\begin{enumerate}
\item \citet{pt05a} collate recent \te\ measurements and use them to
  calibrate simultaneous fits to
  $\rtt\equiv\{f(\otw)+f(\otll)\}/f(\mathrm{H}\beta)$ and a function
  of $\ott \equiv f(\otll)/f(\otw)$.  The latter is a proxy for
  ionization parameter, which is the ratio of ionizing photons to
  hydrogen nuclei present in gas.
\item The photoionization models of \citet{mcgaugh91a} also use both
  \rtt\ and \ott\ as parameters in the diagnostic.  The calibration is
  updated and printed in analytic form by \citet{dmd04a}; we use their
  semi-empirical version.
\item \citet[][hereafter T04]{tremonti04a} use the models of
  \citet[][hereafter CL01]{cl01a} to compute the abundances of
  $\sim$10$^5$ galaxies from the SDSS.  The original models are a
  suite of analytic functions involving different combinations of
  strong emission lines; the choice of diagnostic then depends on the
  spectral data available.  However, \citetalias{tremonti04a} directly
  cross-correlate their data with model spectra to find the best-fit
  abundance.  They then fit a simple analytic function to the upper
  branch of the abundance vs. \rtt\ relation.
\item For comparison with the \citetalias{tremonti04a} \rtt\
  calibration, we include the original \citetalias{cl01a} suite of
  analytic functions.  We compute abundances using the diagnostics of
  Cases A through F, where successive diagnostic cases rely on less
  spectral information than the previous one.  For simplicity, we only
  discuss Cases A and F to represent the range of possibilities in
  spectral information.  The former relies on \ntl/\stl\ and the
  latter on \otl/H$\beta$ (with \ott\ as a weak secondary parameter in
  each case).  We note that the \citetalias{cl01a} diagnostics use
  observed fluxes as inputs, unlike most other diagnostics which use
  fluxes corrected for attenuation by dust.
\item \citet{kd02a} attempt an optimal calibration by comparing their
  photoionization models to previous work and combining several
  diagnostics.  For our galaxies, their `combined' diagnostic reduces
  to the \ntl/\otwl\ diagnostic in almost all cases.
\end{enumerate}

In each case, we discard galaxies that have log($\rtt$)~$\geq1$; such
values are observed in star-forming regions, but they are not common
\citep[e.g.,][]{pt05a}.  In this regime, it is unclear which \rtt\
branch is appropriate and the diagnostics are not well-calibrated.
Furthermore, a very high \rtt\ may be an indication of contributions
to the line emission from processes not related to star formation.
There are nine LIRGs and nine ULIRGs with $\rtt\geq1$.

We also assume the upper branch for \rtt-based diagnostics
\citep{ep84a}.  We use the \nt/\otw\ flux ratio as a branch indicator,
where all galaxies with $f(\nt)/f(\otw) > -1$ are assumed to lie on
the upper branch (e.g., \citealt{dmd04a}; see also the theoretical
plots of \nt/\otw\ and \rtt\ vs. abundance in \citealt{kd02a}).  Eight
galaxies (five LIRGs and three ULIRGs) fail this test.  Seven of these
also surpass the \rtt\ threshold; we exclude the remaining LIRG from
our analysis, to avoid the problem of stitching together upper and
lower branch calibrations (which exists when two different diagnostics
are put together; e.g., \citealt{salzer05a}).  The number of galaxies
with low values of \nt/\otw\ is interesting, and may suggest an even
larger downward spread in abundance than is found by taking into
account only the upper branch systems.

These two cuts do not strongly impact our analysis.  Once we put aside
galaxies with $z > 0.27$, we have a working sample of 55 LIRGs and 22
ULIRGs.  (We discuss further the $z\sim0.4-0.5$ points in
\S\ref{sec:comp-high-redsh}.)

Different diagnostic/calibration pairs can yield very different
abundances for the same galaxy.  Figure~\ref{fig_zscatt} shows the
abundance of a random sampling of our galaxies using each method
listed above.  The methods are ordered on the horizontal axis by
increasing median abundance.  Over the full sample, the extreme median
values (the \citetalias{cl01a} Case A and \citet{pt05a} calibrations)
differ by a factor of 8 for the LIRGs, and correspond to
$12+\mathrm{log(O/H)}=8.2$ and 9.1, respectively.  The `median of the
medians' is 1.25\zsun.  For the ULIRGs, the extreme values are 8.1 and
9.0, also a spread of a factor of 8, with the median being roughly
1.0\zsun.  However, if the \citet{pt05a} diagnostic (the only one in
our sample based solely on $T_e$ abundances) is removed, the scatter
is dramatically reduced.  The peak-to-peak variation among the five
remaining diagnostics is only a factor of 2.  The median abundances
among these five are 1.6\zsun\ and 1.4\zsun\ for the LIRGs and ULIRGs,
respectively.

In the next subsections we discuss sources of uncertainty.

\subsection{Uncertainties in Abundance Caused by Choice of Diagnostic}
\label{sec:uncert-abund-diag}

\begin{figure*}
  \plottwo{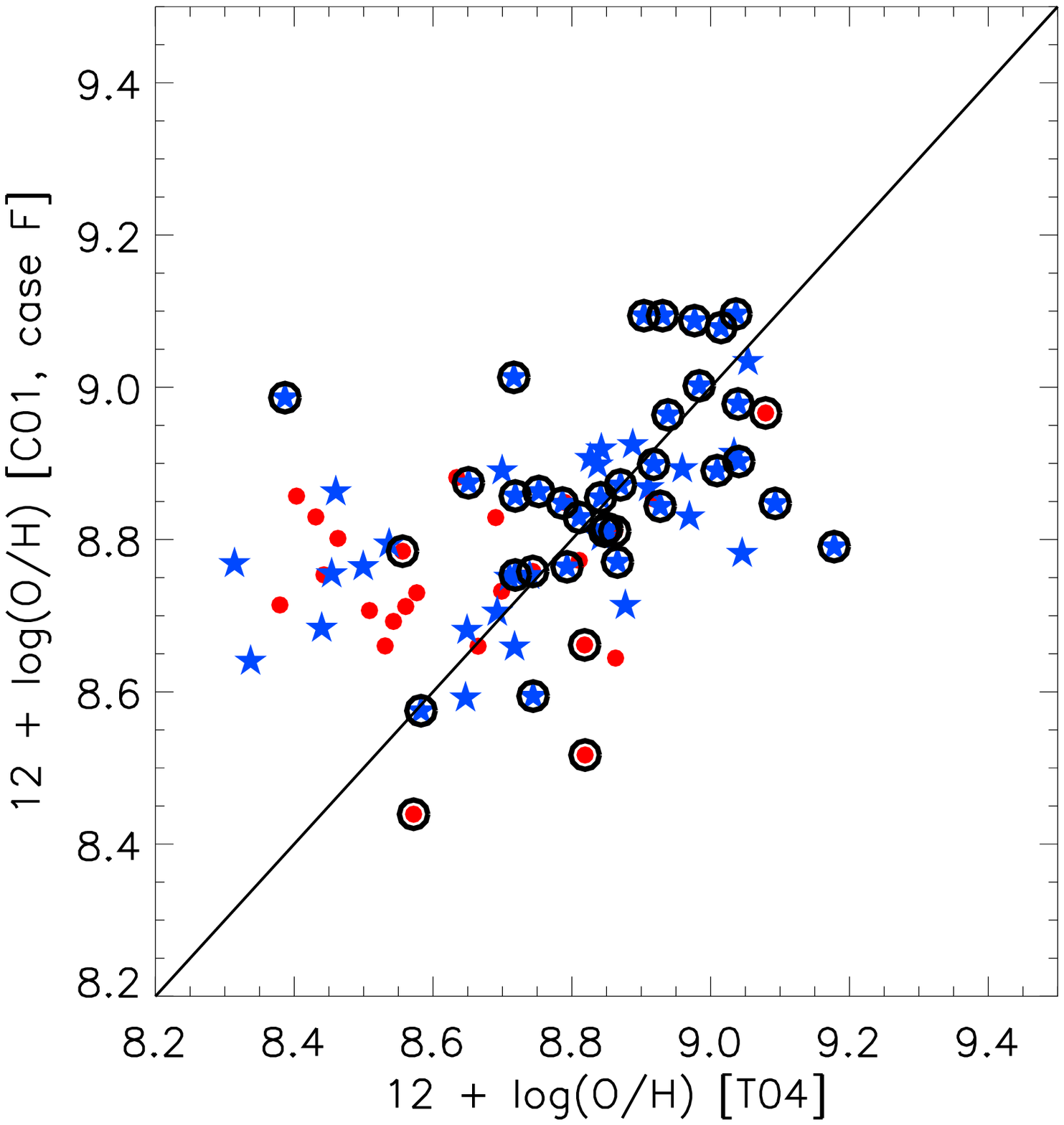}{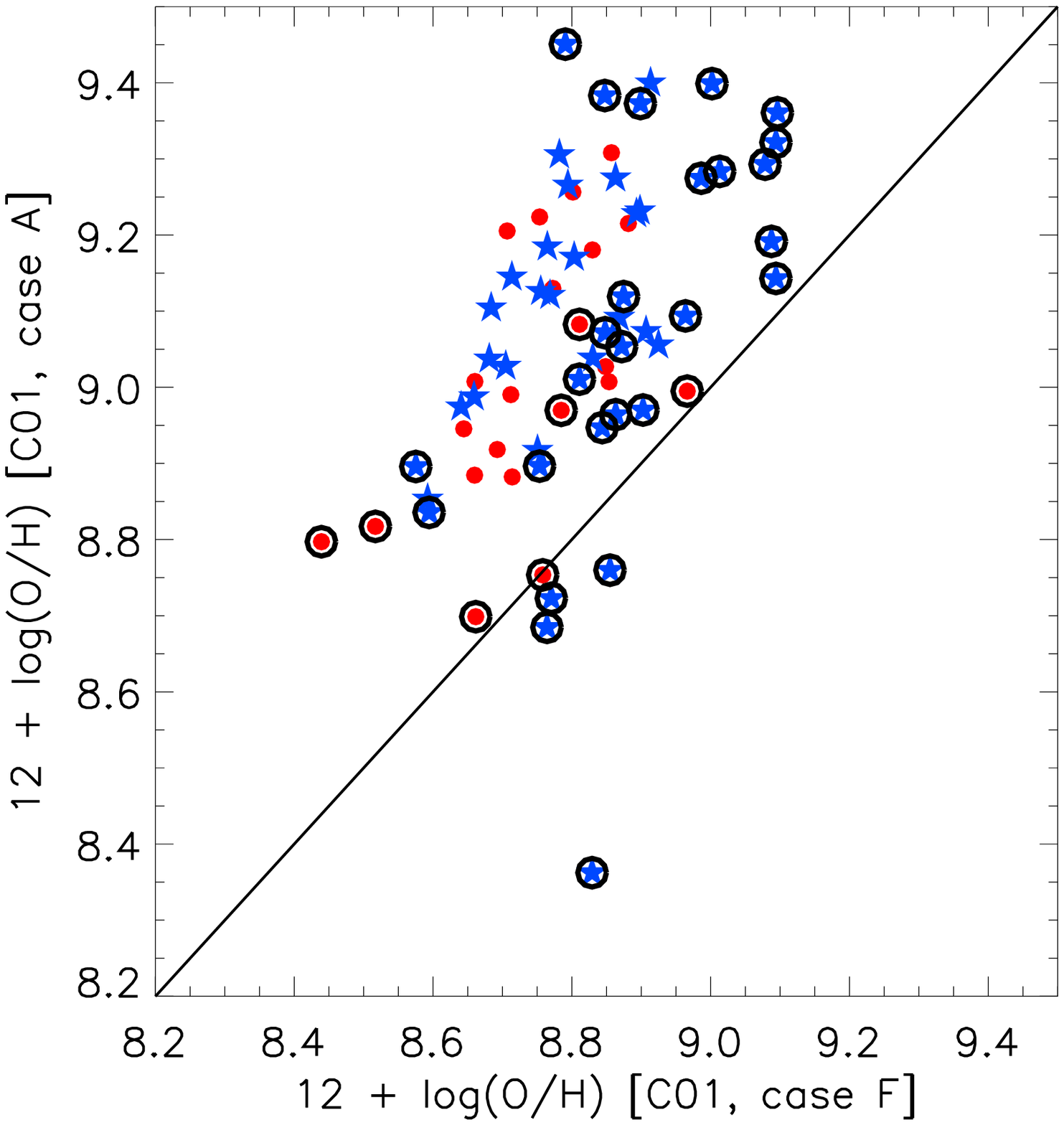}
  \caption{Abundance as a function of calibration for (left)
    \citet{cl01a} Case F vs. \citet{tremonti04a} \rtt\ abundances and
    (right) \citet{cl01a} Case A vs. Case F abundances.  The former
    indicates good correspondence on average except at high \rtt\ (low
    \citealt{tremonti04a} abundances).  The latter illustrates good
    correlation but a significant offset between the two, independent
    of whether or not the galaxies fall inside the SDSS star-forming
    galaxy locus \citep{kauffmann03a}.  Galaxies inside the locus are
    circled.  Galaxy designations follow Figure~\ref{fig_lrats}.}
  \label{fig_zcal}
\end{figure*}

The lowest abundances arise from the use of the empirically-calibrated
diagnostics.  This exemplifies the difference between \te\ abundances
and those from theoretical calibrations.  Discussions of the
weaknesses of each method are found in \citet{kbg03a} and elsewhere,
and we do not expand upon them further.  However, we do note that very
recent measurements of temperature fluctuations in \ion{H}{2} regions
coupled with heavy element recombination line measurements may
alleviate this discrepancy by raising the $T_e$ abundances
\citep{bresolin07a}.  The use of the $T_e$-calibrated \citet{pt05a}
method is thus potentially suspect without an adjustment to agree with
these findings.  Use of their diagnostic calibration is also a
particular challenge for LIRGs and ULIRGs.  We compute low values of
their ionization parameter variable $P$ ($\langle P \rangle = 0.2$)
and relatively high values of \rtt\ ($\langle \mathrm{log}(\rtt)
\rangle = 0.7$ for the LIRGs and 0.9 for the ULIRGs).  These $P$/\rtt\
combinations are not well calibrated for either branch of the \rtt\
vs. abundance diagram in the \citet{pt05a} diagnostic.

Of the photoionization models, those of \citet{kd02a} produce the
lowest dispersion in our sample, as well as the second-highest
abundance.  The low dispersion is due to the fact that the \nt/\otw\
flux ratio is insensitive to the ionization parameter.  One potential
problem with this diagnostic calibration is the form assumed for the
dependence of the N/O abundance ratio on oxygen abundance.
\citet{kd02a} assume that N/O is constant below 0.23\zsun, and linear
in abundance above this value \citep{vsh98a}.  However, the data from
recent studies seem to show that N/O may be driven by a linear
combination of primary and secondary yields, suggesting a sum of the
two contributions rather than an abrupt transition between them.  This
leads to a slightly higher value of N/O for a given abundance
\citep[][and references therein]{bsg05a}.  This should lead to a
higher value of the flux ratio [\ion{N}{2}]/[\ion{O}{2}] for a given
abundance, and thus a lower derived abundance from the calibration.

Perhaps the most important calibration issue for the results of this
paper is the disagreement between the \citetalias{cl01a} Case A$-$F
analytic diagnostics and the \citetalias{tremonti04a} empirical
calibration of \rtt\ based on the \citetalias{cl01a} models.  Rather
than comparing all six \citetalias{cl01a} diagnostics to
\citetalias{tremonti04a}, we use Cases A and F as representative of
the greatest and least amounts of spectral information required by the
\citetalias{cl01a} models.  Figures~\ref{fig_zscatt} and
\ref{fig_zcal} illustrate that, while \citetalias{cl01a} Case F and
\citetalias{tremonti04a} agree reasonably well on average,
\citetalias{cl01a} Case A is higher on average than both by
$\sim$0.3~dex.

What is the cause of this discrepancy?  Either the LIRGs and ULIRGs
have line ratios that are affected by processes unrelated to star
formation (see the next subsection), or the \citetalias{cl01a} Case A
diagnostic does not properly account for these objects.  Looking at
Figures~\ref{fig_lrats} and \ref{fig_zcal}, it is tempting to conclude
that Case A, which like the \citet{kd02a} diagnostic is based on a
line ratio involving \nt, predicts higher abundances due to \nt\
enhancement by a buried AGN or shocks.  Consider the following
scenario: AGN and shocks are important for some galaxies, for which
the different \citetalias{cl01a} diagnostics yield inconsistent
abundances; for purely star-forming galaxies, Cases A and F instead
give the same result.  More specifically, the objects which fall
within the SDSS locus of most star-forming galaxies (a criterion which
securely excludes galaxies with even a small amount of AGN or shock
contribution) should give consistent results for diagnostics A and F
of the \citetalias{cl01a} model, while those outside this locus should
not.  Figure~\ref{fig_zcal} reveals that this is not the case: LIRGs
and ULIRGs which fall beneath the outer SDSS boundary in the
\nt/H$\alpha$ vs. \ot/H$\beta$ line ratio diagram (see
Figure~\ref{fig_lrats}) show almost as large a disagreement between
the \citetalias{cl01a} Cases A and F as do the galaxies that lie
outside it.  Thus, physical effects may have a small contribution, but
it appears that most of the difference is attributable to the
inability of the Case A diagnostic to properly account for the
abundances of these systems.  The details of why this may be so are
beyond the scope of this paper.  As we discuss in the next subsection,
other models are consistent with all of our LIRGs and ULIRGs being
star-forming galaxies.

Among the \citetalias{cl01a} models, the Case F diagnostic minimizes
systematic offsets of LIRGs and ULIRGs with respect to the more
sophisticated method by which \citetalias{tremonti04a} compute
abundances (their method is to find the model spectrum that is the
best fit to the entire observed spectrum, rather than using just one
or two emission-line ratios).  However, as we see in detail in
Figure~\ref{fig_zcal}, the agreement breaks down at the highest values
of \rtt; in this regime, the Case F diagnostic gives higher abundances
by $\sim$0.3~dex.  As Figure~\ref{fig_zscatt} illustrates, this has
little effect on the median abundance.  Our use of
\citetalias{tremonti04a} instead of Case F from \citetalias{cl01a}
thus makes little difference to the results of this paper.  Despite
the fact that it is a single-parameter diagnostic, we choose the
\citetalias{tremonti04a} diagnostic over the Case F diagnostic because
it employs the traditional \rtt\ line ratio and is the empirical
product of a methodologically sophisticated application of the
\citetalias{cl01a} models.

\subsection{Uncertainties in Abundance Caused by Physical Effects}
\label{sec:uncert-abund-phys}

\begin{figure}
  \plotone{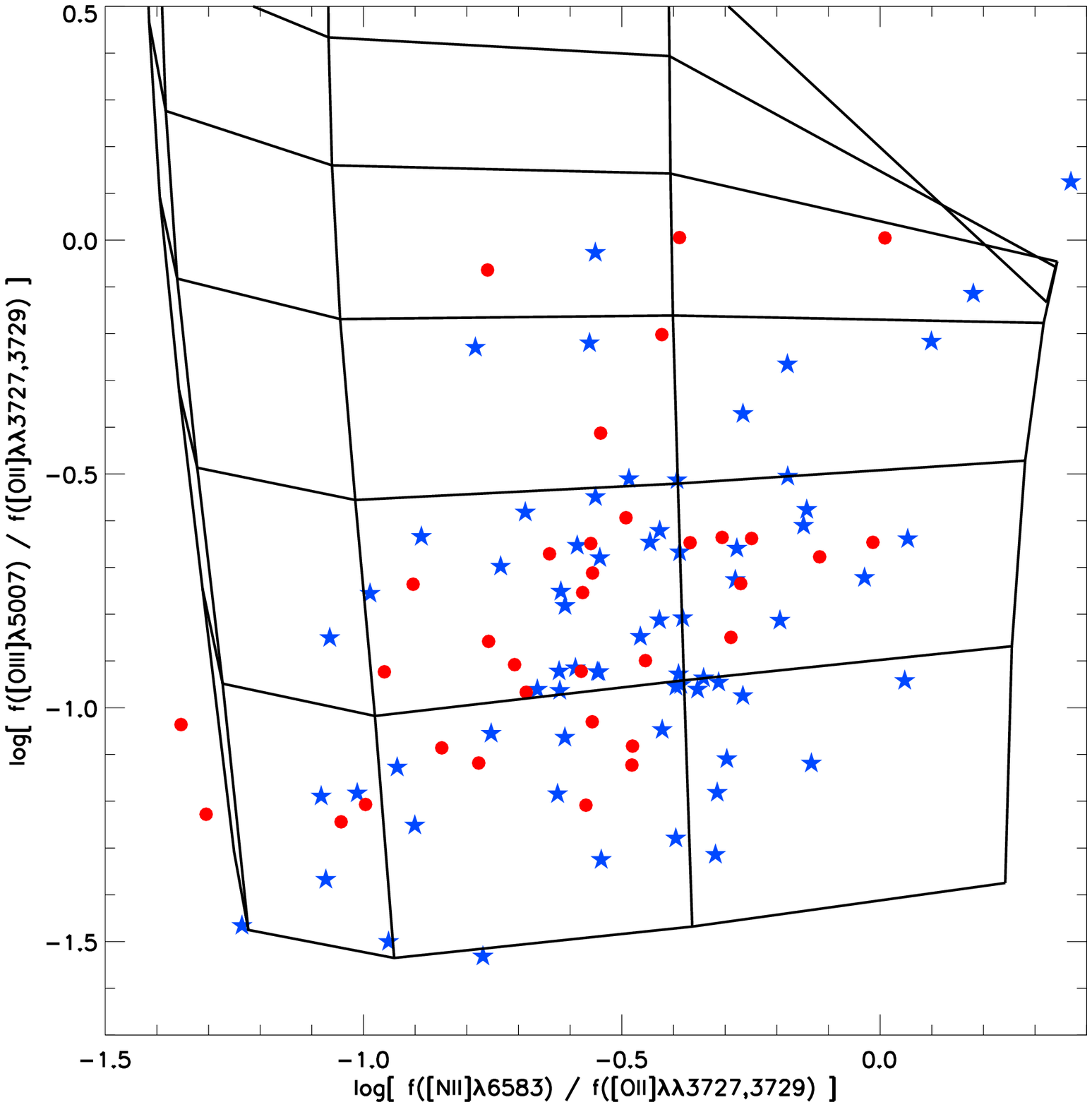}
  \caption{\ntl/\otwl\ vs. \otl/\otw, as measured and predicted by
    theory.  These line ratios nicely separate the effects of
    abundance and ionization parameter, and show that starburst models
    can in principle explain LIRG and ULIRG line ratios.  Abundance
    increases to the right, with grid points of 0.1, 0.4, 0.8, 2.0,
    and 4.0 in units of \zsun.  Ionization parameter increases upward,
    with grid points of (in the log) roughly 1.2 to 3.0, increasing in
    steps of 0.3 dex.  The photoionization models are from
    \citet{kewley01a} and use the Starburst99 stellar synthesis code
    \citet{leitherer99a} with `Lejeune $+$ Schmutz' stellar
    atmospheres as an input.  These models assume an electron density
    of 350~cm$^{-3}$ and a continuous starburst of age 8~Myr.  Galaxy
    designations follow Figure~\ref{fig_lrats}.}
  \label{fig_theory}
\end{figure}

An obvious important consideration is the dust extinction corrections
required to derive intrinsic fluxes. (The optical attenuation derived
from the Balmer decrement and a foreground screen is E(\bv)~$=0.9$ on
average, corresponding to a factor of 3 upward correction to \rtt\ to
reach the unattenuated line ratio.)  LIRGs and ULIRGs have copious
quantities of dust in their nuclei; the emission from this dust
creates their high infrared luminosities.  The optical line emission
we observe does not emerge from the heavily obscured nuclear star
clusters in these systems, where almost all of the star formation is
occurring \citep{sse00a}.  Presumably it arises instead from star
formation occurring near the surface of the nuclear dust cloud, mildly
obscured sight lines, or light scattered into the line of sight.

Our use of the Balmer decrement and assumption of an effective
foreground screen geometry mean that even our `corrected' line fluxes
for these dusty systems will not correspond to their full line
luminosities or star formation rates. However, several considerations
argue that the {\it ratios} of our corrected line fluxes will still
yield fair estimates of abundances.  First, ULIRGs in particular show
such large turbulent velocities within their compact nuclear gas
structures \citep[e.g.,][]{ds98a} that any abundance gradients should
be flattened rapidly once gas has been dumped into a merger's center
of mass.  This means that the outer and inner regions of a merger's
dusty core should have similar abundance values.  Second, our use of
an effective screen geometry is consistent with both (a) the
derivation of the \citet{calzetti00a} attenuation law we have adopted,
and (b) in many objects, the measurement of Balmer decrements much
higher than could be achieved if \htwo\ regions and dust were
uniformly mixed \citep[e.g.,][]{thronson90a}.  Moreover, tests show
that neither use of a Galactic extinction curve \citep{mm72a} nor
assumption of a uniformly mixed geometry (when sources' observed
Balmer decrements allow this) lead to significantly different results.
Finally, we emphasize that any relative underestimate of extinction at
\otwl\ would require that the true \rtt\ be higher, and therefore the
true oxygen abundance lower, than we have reported here.  As
\S\S\,\ref{sec:lzreln} and \,\ref{sec:mzreln} make clear, such an
adjustment would only strengthen this paper's conclusions.

A second relevant issue is the applicability of star-formation
calibrated abundance diagnostics.  As we discuss in
\S\ref{sec:sample-selection}, LIRGs and ULIRGs deviate from the main
body of star-forming galaxies in many line-ratio diagrams (Figure
\ref{fig_lrats}).  However, they generally lie completely below the
\citet{kewley01a} maximal starburst lines in at least two of these
diagrams, and so many or all can plausibly be treated as starbursts.

We illustrate this in Figure~\ref{fig_theory} by plotting the
\otl/\otw\ line ratio (note this is $0.75 \times \ott$) as a function
of \nt/\otw\ for each galaxy.  Photoionization models from
\citet{kewley01a} are plotted atop the data, assuming a continuous
starburst of age 8~Myr.  These line ratios nicely separate the effects
of abundance and ionization parameter.  LIRGs and ULIRGs in this
diagram have relatively high abundances and low ionization parameter,
though the details depend somewhat on the adopted starburst model.

It remains possible that some objects in our sample may have emission
line fluxes with a non-zero contribution from gas that has been
photoionized by an AGN or collisionally ionized by shocks.  The
addition of either a low-luminosity or high-luminosity AGN moves
galaxies away from the starburst locus
\citep[e.g.,][]{vo87a,kewley06a}.  However, it appears that AGN do not
make a significant contribution, because mid-infrared diagnostics of
ULIRGs that distinguish starbursts and AGN agree with optical
diagnostics if all non-Seyfert ULIRGs are assumed to be
starburst-powered \citep{lvg99a}.  The addition of shocks also moves
line ratios away from the region of pure starburst galaxies
\citep{ds95a}.  The existence of shocks would be consistent with the
presence of massive, fast, galaxy-scale winds in starburst-dominated
LIRGs and ULIRGs
\citep{heckman00a,rvs02a,rvs05a,rvs05b,martin05a,martin06a}.
\citet{rvs05b} also find evidence that ULIRGs with LINER spectra have
higher wind velocities than those with \htwo\ spectral types,
suggesting that wind shocks could play a role in exciting the observed
optical emission lines in LINERs.

Without careful modeling, we cannot easily flag galaxies with
significant contributions from low-level AGN or shocks.  Rather than
attempt an extensive comparison to models of stellar and AGN
photoionization or shocks, we make three empirically and theoretically
motivated cuts in our data, resulting in three partially overlapping
subsamples.  Computing our results for each subsample allows us to
ascertain the effect of including galaxies with possible low-level AGN
or shock contamination.  These cuts are as follows: (1)~accepting only
galaxies which lie beneath the maximal starburst line
\citep{kewley01a} in the \ot/H$\beta$ vs.  \nt/H$\alpha$ and
\ot/H$\beta$ vs. (\stl)/H$\alpha$ diagrams; (2)~accepting only
galaxies which lie beneath the maximal starburst line
\citep{kewley01a} in the \ot/H$\beta$ vs.  \ool/H$\alpha$ diagram; and
(3) accepting only galaxies which lie beneath the line denoting the
boundary of the locus of SDSS star-forming galaxies
\citep{kauffmann03a}.  As shown in later sections, we find that our
results are not significantly altered by picking one subsample over
another.

A final uncertainty relates to the apertures covered by the
spectroscopic observations.  Disk galaxies have radial abundance
gradients, in the sense that abundance decreases with increasing
radius \citep[e.g.,][]{zkh94a}.  Large apertures covering a
significant fraction of the galaxy's light thus lead to lower computed
abundances than those covering only the galaxy's inner regions.  For
the Nearby Field Galaxy Sample (NFGS), covering a range of Hubble
types, the average difference between nuclear and integrated
abundances is $\sim$0.2~dex for nuclear apertures of 1.5~kpc or less
\citep{kjg05a}.

Our spectra include data with a variety of physical apertures, since
they are produced with a variety of slit widths and extraction
apertures.  Our LIRG and ULIRG samples have median redshifts of 0.04
and 0.14, respectively.  The LIRG data in particular include
one-dimensional spectroscopic apertures $\sim1\arcsec-3\arcsec$ in
width, which corresponds to physical scales of $\sim1-2$~kpc.  These
qualify more or less as nuclear spectra according to the
\citet{kjg05a} convention for the NFGS.  The ULIRG apertures also
range from $\sim1\arcsec-3\arcsec$ in diameter, with about half having
small apertures.  However, being at higher redshift, they subtend
apertures ranging from $2-7$~kpc.  Based on comparison to local field
galaxies, we would then expect a slight systemic deviation of the
average ULIRG to lower abundances relative to the average LIRG (by
$\sim$0.1~dex; \citealt{kjg05a}).  However, the existence of radial
abundance gradients in evolved mergers may be inconsistent with
expectations of strong radial mixing of heavy elements (L. Kewley
2007, private communication; \citealt{eg95a}).  If there has been strong
radial mixing, aperture size will have little affect on the integrated
abundance.

In this work we use two local comparison samples, as well as high
redshift data.  We defer discussions of aperture effects related to
these samples to their points of introduction later in the paper.

With these caveats in mind, we proceed in the next sections to employ
the internal consistency of a given abundance diagnostic/calibration
pair (the \citetalias{tremonti04a} \rtt\ calibration in our case) to
compare LIRGs and ULIRGs to other galaxy populations.

\section{NEAR-INFRARED LUMINOSITY-METALLICITY RELATION}
\label{sec:lzreln}

The best way to interpret the abundances of local LIRGs and ULIRGs in
the context of galaxy evolution is to compare to the
luminosity-metallicity (\lz) and mass-metallicity (\mz) relationships
of other star-forming galaxies (i.e., galaxies of lower star formation
rate or those selected at other wavelengths).  These describe the
increase of abundance with increasing galaxy luminosity and mass.
They reflect, for a given gas mass fraction, either (a) a sequence of
changing star formation history and enrichment with stellar mass or
(b) preferential accretion of under-enriched gas or loss of
over-enriched gas in low-mass galaxies.  Recent authors have argued
for both scenarios (\citetalias{tremonti04a}; \citealt{brooks07a}; but
see also \citealt{kwk06a} for discussion of an origin in a variable
initial mass function).  We start with the \lz\ relation, since
luminosity is a more easily measured quantity than mass.

We compare to the only available near-infrared $\lz$ relation for
starbursts \citep{salzer05a}.  The advantage of using observations in
the near-infrared is their lower sensitivity to dust and mass-to-light
ratio variations than optical observations \citep{bd00a,salzer05a}.
The galaxies comprising this study are a subsample of the KPNO
International Spectroscopic Survey \citep[KISS;][]{salzer00a}, which
is an objective prism survey of emission-line galaxies in the local
universe.  The KISS subsample from \citet{salzer05a} consists of
galaxies with near-infrared photometry, mostly taken from 2MASS.  The
galaxies have $\langle z \rangle = 0.063$ and $K_s$-band absolute
magnitudes ranging from -16.5 up to -25 (a factor of 3000 in
luminosity).

Because of the low redshifts and small spectroscopic apertures
($1.5\arcsec-2\arcsec$, corresponding to 2~kpc) for this sample, the
computed abundances are approximately nuclear, and we make no aperture
corrections.

\subsection{NIR photometry} \label{sec:nir-photometry}

Near-infrared photometry for our LIRG and ULIRG sample is taken from
ground-based data in the \kp\ band where available
\citep{sse00a,stanford00a,kvs02a}, and from the 2MASS Large Galaxy
Atlas \citep{jarrett03a} or Extended Source Catalog \citep{cutri06a}
otherwise.  We use `total,' or extrapolated, $K_s$ magnitudes from the
2MASS catalog.  For some objects with multiple nuclei, a 2MASS
magnitude is not available for the nucleus of interest but the sky-
and star-subtracted image is available \citep{cutri06a}.  In these
cases, we used the 2MASS Atlas images (in `postage-stamp' form) to do
aperture photometry on the nucleus, including associated diffuse
emission.  In other instances a second nucleus was already removed
through star subtraction, which was to our advantage.  However, in the
few instances where this subtraction was poor we did the photometric
separation by hand.  For the ULIRGs with published nuclear and total
magnitudes \citep{kvs02a}, we used the relative nuclear magnitudes to
scale the total magnitude for the nucleus of interest.  Finally, we
note that the published FIRST-FSC magnitudes are on average only 70\%
of the observed total, as discussed in \citet{stanford00a}; in these
cases we made an average upward correction to reach $\sim$100\%.

We converted measured \kp\ magnitudes to $K_s$ magnitudes using the
formula from \citet{wc92a}; the average \hkp\ color for \htwo\ ULIRGs
from \citet{sse00a} ($\langle \hkp \rangle \sim 0.5$, which applies
roughly to LIRGs as well; see \citealt{scoville00a}); and the color
transformation from CIT system $K$ magnitudes to $K_s$ magnitudes
\citep{cutri06a}.

Starting with the total absolute magnitudes, we contemplated a number
of adjustments to the measured luminosities to reach the rest-frame
host galaxy luminosity.  We decided to apply only one of these
adjustments.  This involves removal of the central point source in
ULIRGs, which is most likely an extremely luminous star cluster (but
may also contain part of the central bulge; \citealt{sse00a}).  Very
accurate point-source subtraction is allowed by high-resolution {\it
  Hubble Space Telescope} ({\it HST}) near-infrared data of ULIRGs
\citep{veilleux06a}.  The data for five \htwo\ galaxies from this
sample have an average point-spread function (PSF) to host-galaxy
(PSF-subtracted) luminosity ratio of 0.07, with a range of 0.01 to
0.13.  Applying this average correction to each ULIRG lowers its
luminosity by an insignificant 0.08 magnitudes.

We also contemplated corrections for dust extinction/emission.  Most
of the near-infrared light in LIRGs and ULIRGs that are dominated by
star formation arises in modestly extinguished stars, not hot dust.
For cold, star-forming LIRGs and ULIRGs, most of the near-infrared
emission is extra-nuclear \citep{carico90a,scoville00a,veilleux06a},
and the colors of this extended light are consistent with those of
normal spirals in the case of LIRGs and a reddened stellar population
in the case of ULIRGs \citep{carico90a,scoville00a,dbw02a}.  The
average predicted screen extinction is modest even for the more
heavily extinguished nuclear regions ($A_V$ of a few;
\citealt{scoville00a,sse00a,dbw02a}).  To achieve an unreddened
luminosity we would correct upward by only a few tenths of a magnitude
\citep{draine03a}, and even less for the more dominant extended
emission.  Modest hot dust emission may also contribute to the nuclear
light \citep{sse00a,dbw02a}, but neither the dust emission nor nuclear
light is significant.  As a final piece of evidence, ULIRGs tend to
show elliptical-like global surface-brightness profiles in the
near-infrared, typical of relaxed stellar populations \citep[][and
references therein]{vks02a}.

Finally, we considered K-corrections.  There is a strong diversity in
observed spectral energy distribution (SED) shapes in ULIRGs, but the
average ULIRG shows increasing $\nu L_\nu$ from optical to infrared
wavelengths, and flat or increasing values in the near-infrared
\citep{tks99a,farrah03a}.  These yield positive K-corrections in the
$K$-band that increase slowly with increasing redshift and depend
somewhat on the actual SED shape ($0.06-0.25$~mag at $z\sim0.3$ and
$0.1-0.4$~mag at $z\sim0.6$; \citealt{tks99a}).  Given the similarity
of the near-infrared colors of LIRGs and ULIRGs (see previous
paragraph and references therein), K-corrections for LIRGs are of
similar magnitude.

We conclude that the K-corrections, extinction corrections, and
corrections for hot dust emission are on average small (a few tenths
of a magnitude or less in each case).  Given the inherent variations
in SEDs and our inability to determine precise adjustments for each
galaxy, we choose to ignore these corrections.  Regardless, their
effect on the following results is negligible.  If anything, they are
likely to increase the significance of the observed effects.

\subsection{L-Z relation} \label{sec:l-z-relation}

\begin{figure}
  \plotone{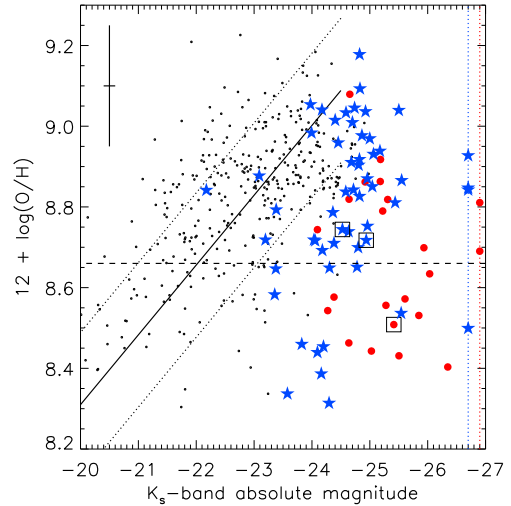}
  \caption{$K_s$ band luminosity-metallicity relation for nearby
    emission-line galaxies (small black open circles), LIRGs (blue
    stars), and ULIRGs (red filled circles).  Most of the LIRGs and
    ULIRGs fall well below the \lz\ relation.  The nearby galaxies are
    from the KISS sample, and the black line and dotted lines are a
    fit to the data and 1$\sigma$ RMS dispersion, respectively
    \citep{salzer05a}.  The dashed line locates solar abundance.
    Boxed points do not pass our first emission-line cut.  We do not
    plot points with $z > 0.3$.  The far-right points do not have
    measured K magnitudes.}
  \label{fig_lz1}
\end{figure}

\begin{figure}
  \plotone{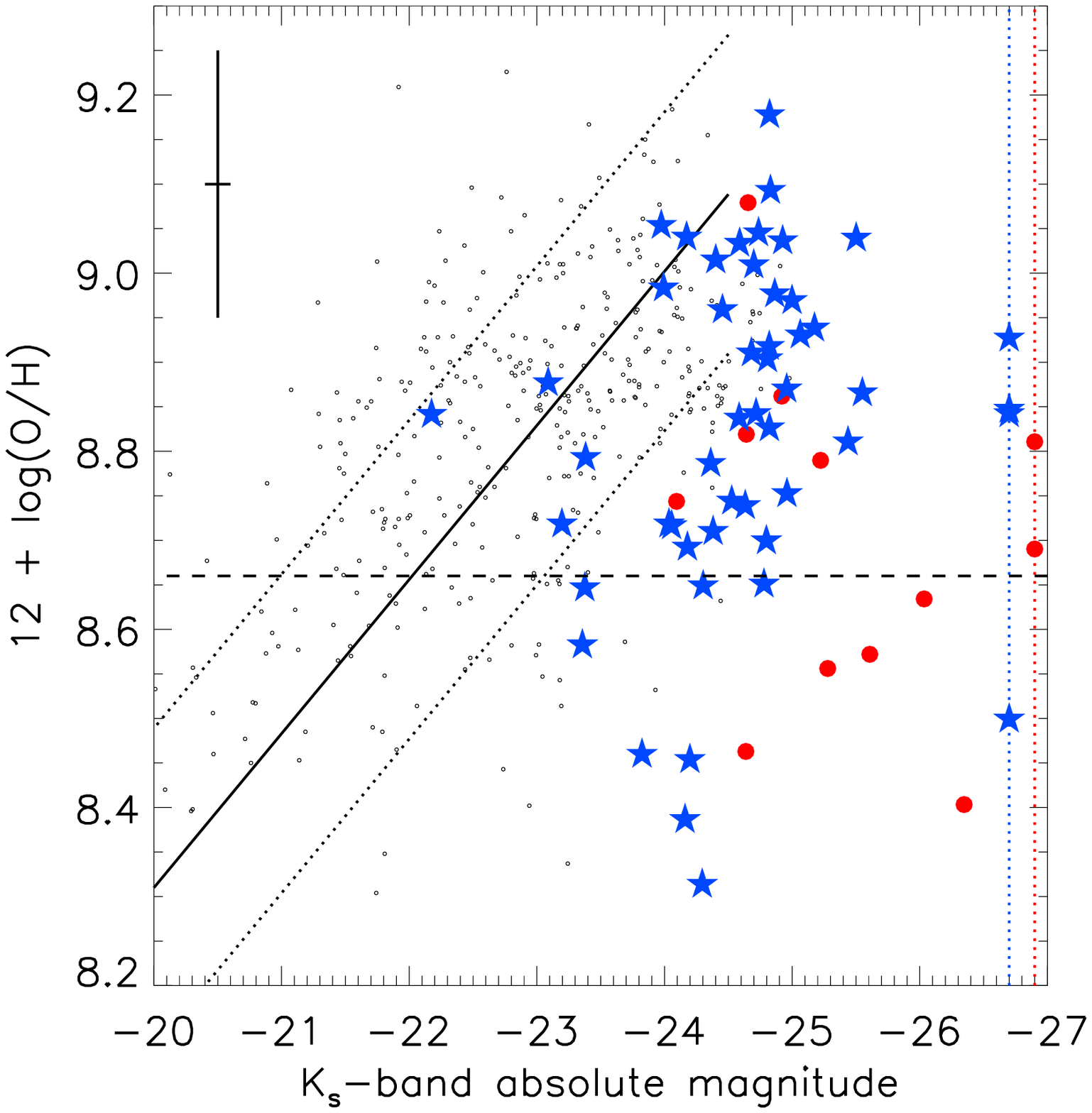}
  \caption{\lz\ relation, but only for LIRGs and ULIRGs that pass our
    second emission-line cut.  See Figure~\ref{fig_lz1} for more
    details.}
  \label{fig_lz2}
\end{figure}

\begin{figure}
  \plotone{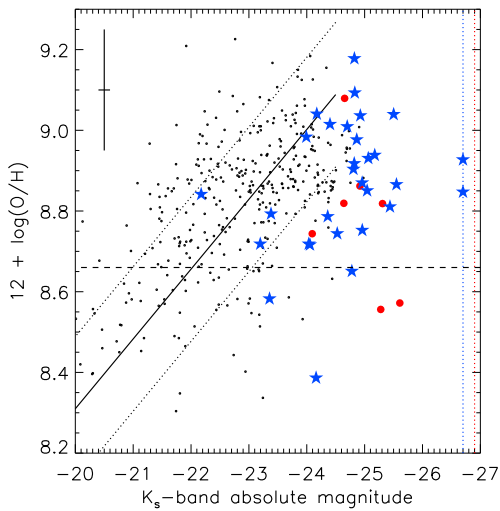}
  \caption{\lz\ relation, but only for LIRGs and ULIRGs that pass our
    third emission-line cut.  See Figure~\ref{fig_lz1} for more
    details.}
  \label{fig_lz3}
\end{figure}

We use the \citetalias{tremonti04a} \rtt\ abundance calibration to
compute upper branch abundances for our LIRGs and ULIRGs for
comparison to the KISS results.  The resulting $K_s$-band \lz\
relation is shown in Figures $\ref{fig_lz1} - \ref{fig_lz3}$.  Each
figure represents the \lz\ relation using a different emission-line
cut, as described in \S\ref{sec:uncert-abund-phys}.  It is immediately
clear that many LIRGs and ULIRGs do {\it not} fall on this relation,
even if we discard possibly shock-excited or AGN-contaminated sources.
There is some overlap, however.  We also see that the scatter in the
computed LIRG and ULIRG abundances for a given magnitude is higher
than in the KISS comparison sample.

The observed offset is one of abundance and/or luminosity.  If the
LIRGs and ULIRGs are over-luminous in the near-infrared with respect
to galaxies of lower star formation rates, then there must be a
significant (factor of $\sim$10) increase in the near-infrared
emission above that from the old, mass-tracing stellar population.  As
we discuss in \S\ref{sec:nir-photometry}, hot dust could increase the
luminosity above that expected from the stellar population, but only
by a small fraction of the total.  Massive red supergiants from very
young star forming regions can also add to the luminosity from old
stars while keeping the near-infrared colors almost constant
\citep{leitherer99a}.  However, the heaviest young star formation is
nuclear, and contributes only a small percentage of the total
near-infrared light (\S\ref{sec:nir-photometry}).

We conclude that hot dust and young stars cannot come close to
producing a factor of 10 increase in luminosity.  Thus, LIRGs and
ULIRGs are under-abundant with respect to local,
emission-line-selected star-forming galaxies of lower star formation
rate.  This conclusion is solidified by comparing LIRGs and ULIRGs to
the local mass-metallicity relation in \S\ref{sec:mzreln}.

The average galaxy in our sample is more NIR-luminous than the range
of luminosities probed by the KISS data (which are biased toward
low-luminosity objects because of the required emission-line contrast;
\citealt{salzer05a}).  Thus, there is uncertainty in the shape of the
\mz\ relation at the luminosities probed by our sample.  However, even
if the relation is conservatively assumed to flatten above $M_{K_s}
\sim -24.5$ rather than to continue to increase, almost all of the
observed galaxies are under-abundant compared to the mean oxygen
abundance of local, emission-line, modestly star-forming galaxies.

\begin{figure}
  \plotone{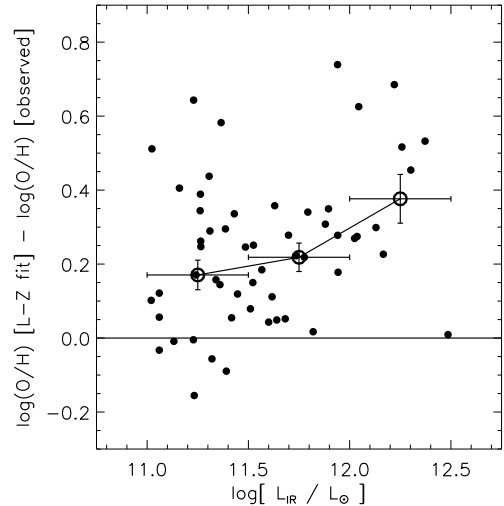}
  \caption{Difference between the observed abundances in LIRGs and
    ULIRGs and the \lz\ relation as a function of infrared luminosity.
    The small filled circles represent individual deviations from the
    \lz\ relation (from the second emission-line cut).  The thick open
    circles are median deviations from \lz\ for equal-size bins
    centered on log(\lir/\lsun) $=11.25$, 11.75, and 12.25 and the
    error bars represent the standard error in the mean in each bin.
    LIRGs are offset by $0.2$~dex, and ULIRGs by $0.4$~dex.
    Comparison to the \lz\ relation shows a mildly significant trend
    toward higher abundance offsets for higher \lir.}
  \label{fig_dzlzlir}
\end{figure}

Figure~\ref{fig_dzlzlir} illustrates the offset in abundance from the
emission-line galaxy relation as a function of total infrared
luminosity \lir\ for the galaxies passing the second emission-line
cut, and conservatively assuming a flat relation above $M_{K_s} \sim
-24.5$.  Plotting the data in three bins of equal width in log(\lir)
shows qualitatively that galaxies of higher \lir\ have higher offsets
from \lz, with the LIRGs offset by 0.2~dex and the ULIRGs by 0.4~dex.
Quantitatively, there is a weak and marginally significant correlation
between the two, with the offset increasing by $\sim$0.2~dex for each
1~dex increase in \lir.  The correlation coefficient is 0.3 regardless
of the emission-line cut chosen, while the (parametric) significance
of the correlation is 99.7\%, 98\%, and $\sim$90\% for the first,
second, and third cuts, respectively, decreasing largely due to the
decreasing number of points in each cut.  The significance of the
correlation depends also on the NIR luminosity at which the flattening
is assumed to occur: it increases (decreases) as the cutoff NIR
luminosity increases (decreases).  It also depends on the
emission-line calibration chosen; some calibrations give smaller
LIRG-to-ULIRG discrepancies (Figure~\ref{fig_zscatt}).  Finally, we
caution that the trends may be weaker than suspected from the \lz\
relation alone (see \S\ref{sec:mzreln}).

\begin{figure*}
  \plottwo{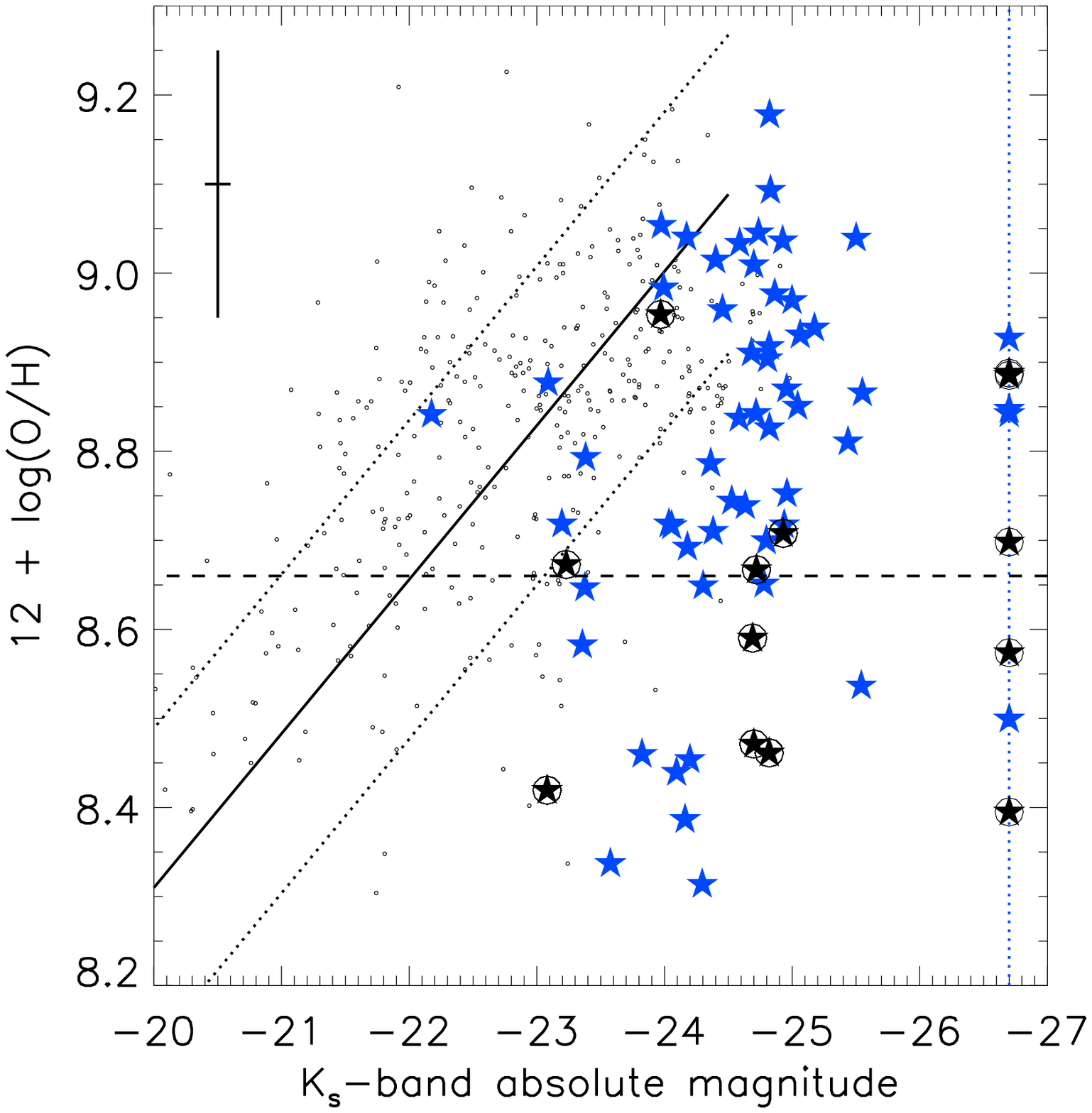}{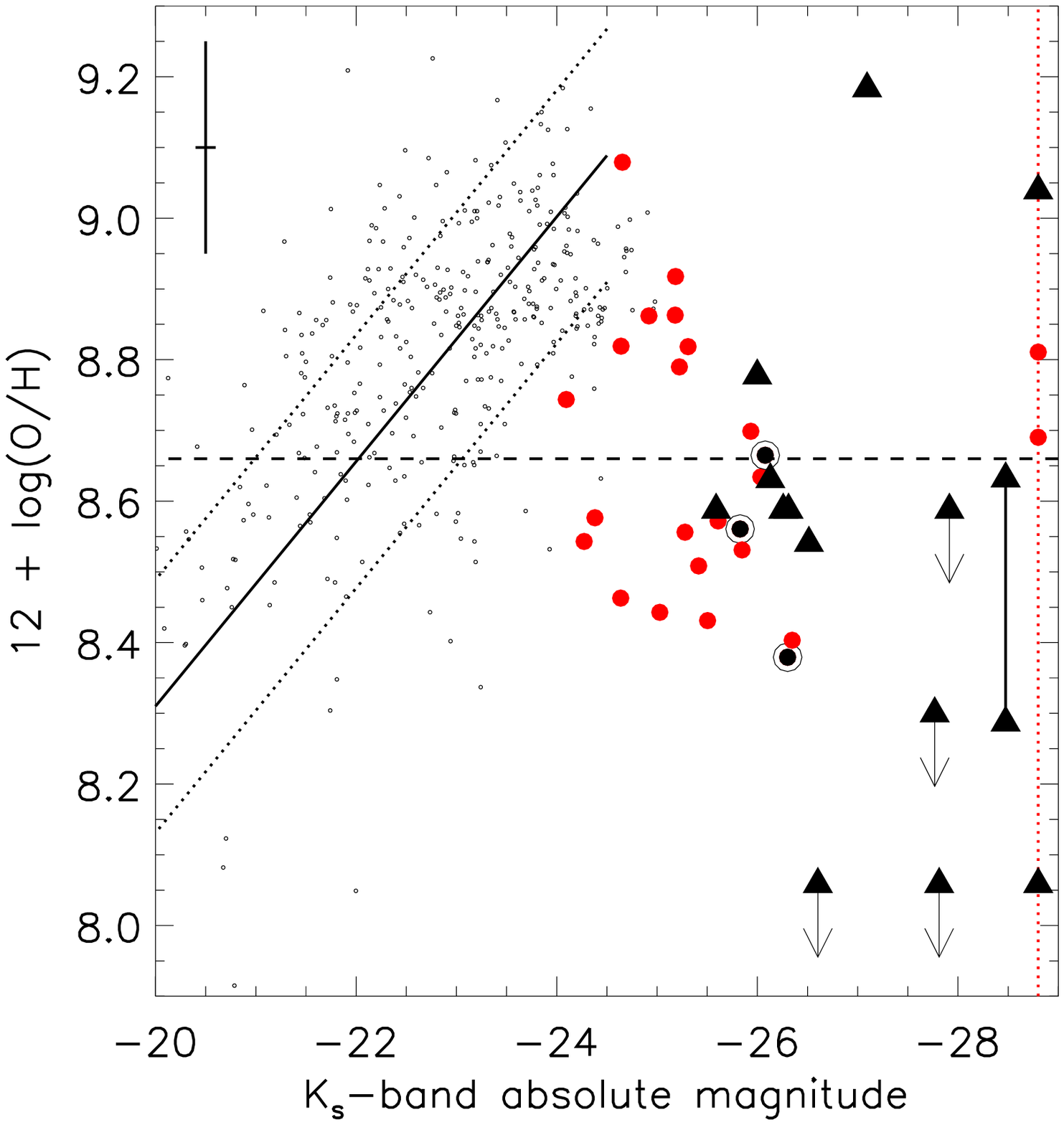}
  \caption{Redshift evolution of the LIRG (left) and ULIRG (right) L-Z
    relations.  See Figure~\ref{fig_lz1} for more details.  We show
    here data that pass the first emission-line cut.  (a) The black,
    encircled stars are LIRGs with $0.4 < z < 0.9$ from
    \citet{liang04a,lhf06a}, supplemented with one point from this
    work.  (b) The black, encircled circles are ULIRGs with
    $z\sim0.4-0.5$ from this work.  The black triangles are
    submillimeter-selected galaxies with $1.4 < z < 2.7$ and
    luminosities equal to or greater than those of ULIRGs
    \citep{smail04a,swinbank04a}.  The vertical line connects the
    abundances for the two nuclei of SMM J14011$+$0252
    \citep{nesvadba07a}.}
  \label{fig_lzhiz}
\end{figure*}

\subsection{Comparison To Other Merging Systems}
\label{sec:comp-other-merg}

LIRG and ULIRG under-abundances are consistent with a recent study of
optically-selected mergers \citep{kewley06b}.  These authors compute
the nuclear abundances of a sample of nearby interacting field
galaxies with projected separations of $4-60$~kpc and optical
luminosities $-22 \leq M_R \leq -19$.  They find an under-abundance in
mergers with respect to the optical $B$-band \lz\ relation for
isolated field galaxies.  There is a dependence of the offset on
separation and `central burst strength,' in the sense that galaxy
pairs with small separations ($4-20$~kpc) and strong starbursts also
have the lowest abundances on average compared to isolated galaxies.
They also find that more optically-luminous galaxies have a smaller
offset in abundance; for separations of $4-20$~kpc and $M_B \ga -21$,
the abundance offset is $\la$0.1~dex.

Local LIRGs and ULIRGs have mean optical luminosities comparable to
the highest-luminosity galaxies in the \citet{kewley06b} sample, with
$M_R \sim -22$ \citep{vks02a,ishida04a}.  Based on their nuclear
separations \citep{ishida04a}, the non-isolated LIRGs may represent
similar galaxy interaction states compared to the optically-selected
galaxies.  The mean ratio of present-to-past star formation is higher
in LIRGs, however, based on the higher H$\alpha$ equivalent widths in
LIRGs \citep{veilleux95a,bgk00a}.  The ULIRGs represent stronger or
more highly progressed galaxy interactions with much higher star
formation rates compared to the optically-selected mergers.  This
conclusion is based on the structural properties of ULIRGs
\citep{vks02a} and their high H$\alpha$ equivalent widths
\citep{vks99a}.  More indirectly, ULIRGs are scarce in the local
universe relative to the parent sample of the optically-selected pairs
\citep{falco99a}.

The observed differences between LIRGs and ULIRGs and isolated
galaxies of lower star formation rate reflect a continuation of the
trends observed by \citet{kewley06b}.  LIRGs and ULIRGs are more
strongly offset from the $K$-band \lz\ relation (by $0.2-0.4$~dex)
than are galaxies of similar $R$-band luminosity from the $B$-band
\lz\ relation ($\la$0.1~dex; \citealt{kewley06b}).  As we describe
above, they also are stronger or later stage interactions and/or have
stronger starbursts.

Within our sample, the apparent trend of greater abundance offset with
increasing infrared luminosity (Figure~\ref{fig_dzlzlir}) mirrors the
correlation of abundance offset with starburst strength found in
optically-selected mergers.  We also studied the dependence of offset
on projected nuclear separation.  No significant relationship was
found, unlike the optical sample.  This most likely reflects either
(a) the saturation of the sensitivity of nuclear separation to
interaction timescale or merger age as two nuclei merge and/or (b) a
greater correlation of \lz\ offset with level of star formation than
with nuclear interaction state.

\subsection{Comparison To High Redshift} \label{sec:comp-high-redsh}

Comparison of luminosity-metallicity and mass-metallicity relations at
low and high redshift point to chemical evolution of galaxies over
cosmic time
\citep{lcs03a,kk04a,maier05a,savaglio05a,erb06a,mouhcine06a}.  As
mentioned in \S\ref{sec:introduction}, high-redshift abundance
measurements of a modest number of LIRGs selected at 15~\micron\ and
ULIRGs selected at submillimeter wavelengths exist
\citep{liang04a,swinbank04a,tecza04a,nesvadba07a}.  Three of these
studies are based on \rtt\ abundances, while the third
\citep{swinbank04a} relies on the \nt/H$\alpha$ flux ratio, a slightly
less robust indicator.  From our work, we also have data on four LIRGs
and ULIRGs at $z \sim 0.4-0.5$ \citep{rvs05a}.  Because near-infrared
photometry for many of these sources exist
\citep{lhf06a,hammer97a,smail04a}, we can compare them directly to our
data on the $K$-band \lz\ relation.

Of the 15~\micron\ sources from \citet{liang04a}, we select only those
which have infrared luminosities consistent with the definition of
LIRGs.  We assume the $K$-correction is small and do not apply it.
However, we do adjust the magnitudes listed in \citet{lhf06a}
downwards by 1.9 magnitudes to convert from the AB to the Vega
photometric system, as they are listed as AB magnitudes in
\citet{hammer97a}.  (The magnitude of Vega in the AB system is taken
from \citealt{tv05a}.)  We also apply upward aperture corrections of
0.1~dex to reach nuclear abundances, following the treatment of local
disk galaxies in \citet{kjg05a}.

All of the submillimeter galaxies (SMGs) in \citet{swinbank04a} have
ULIRG-like or higher total infrared luminosities.  To compare them to
the local near-infrared \lz\ relation, we apply K-corrections to the
$K$-band observed magnitudes from \citet{smail04a} using the average
$z = 2$ ULIRG-derived correction from \citet{tks99a} of 0.45.  We also
discard galaxies with very broad H$\alpha$ (FWHM~$> 1500$~\kms) and/or
log(\nt/H$\alpha$) $>-0.1$.  This is an attempt to eliminate galaxies
with AGN and/or with emission lines contaminated by shock excitation,
respectively.

Because of limited emission-line data, we compute SMG abundances using
the `coarse' calibration from \citet{salzer05a} that employs
\ntl/H$\alpha$ (and is derived from the \citetalias{tremonti04a} \rtt\
relation).  The exception is SMM~J14011$+$0252, for which we use the
\rtt\ diagnostic with the data from \citet{nesvadba07a}.  Comparison
of the \nt/H$\alpha$ and \rtt\ diagnostics for local ULIRGs suggest
that at low abundances the former tends to over-predict the abundance
by an average factor of $\sim$2.  This is unsurprising given the
results of the application of other \nt-based diagnostics to LIRGs and
ULIRGs (\S\ref{sec:uncert-abund-diag}).

\begin{figure}
  \plotone{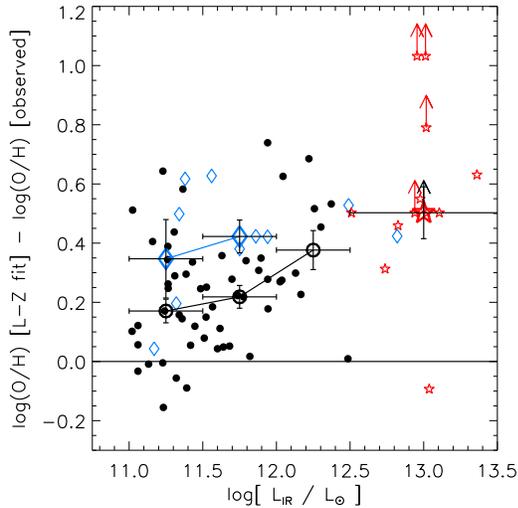}
  \caption{The abundance offset from the local \lz\ relation for low-
    and high-redshift LIRGs and ULIRGs, as a function of infrared
    luminosity.  We show here data that pass the second emission-line
    cut.  The (black) small filled circles, (blue) small open
    diamonds, and (red) small open stars represent $z\sim0.1$ LIRGs
    and ULIRGs from this work, $z\sim0.6$ LIRGs and ULIRGs from
    \citet{liang04a} and this work, and $z\sim2$ SMGs
    \citep{swinbank04a,nesvadba07a}, respectively.  The (black) thick
    open circles, (blue) thick open diamonds, and (red) thick open
    star are median deviations from \lz\ for local LIRGs and ULIRGs,
    $0.4 < z < 0.9$ LIRGs, and $z\sim2$ SMGs, respectively.  The LIRGs
    clearly evolve upward in abundance by $\sim$0.2~dex from
    $z\sim0.6$ to $z\sim0.1$, as would be expected from continual
    processing of heavy elements.  Though there only 2 $z\sim0.5$
    ULIRGs in this figure, there is also apparent redshift evolution
    in ULIRG abundance from $z\sim0.5$ to $z\sim0.1$.  Finally, there
    is also evidence for modest evolution from $z\sim2$ SMGs to
    $z\sim0.1$ ULIRGs, though the observed scatter and systematic
    uncertainties are large.}
  \label{fig_dzlzlirhiz}
\end{figure}

In Figure~\ref{fig_lzhiz}, we plot the \lz\ data for local LIRGs and
ULIRGs, for $z=0.4-0.9$ LIRGs and ULIRGs, and for $z\sim2$ SMGs.  For
SMM~J14011$+$0252, we plot the abundances of both nuclei, J1n and J1s.
Figure~\ref{fig_dzlzlirhiz} makes this comparison more quantitative.
The results show that the LIRGs evolve upward in abundance by
$\sim$0.2~dex from $z\sim0.6$ to the present day, as would be expected
from continual processing of heavy elements in their progenitors.  For
the $z\sim0.5$ ULIRGs, we are limited by number statistics, but there
is a suggestion of evolution.

The very high redshift systems, the SMGs at $z \sim 2$, show a large
scatter in abundances.  This is partly due to low emission-line
sensitivity at these redshifts, the coarse diagnostic required to
compute the abundance, and the difficulty in weeding out AGN or
shock-excited candidates.  Furthermore, the SMGs are much more near-
and far-infrared luminous than ULIRGs, so the comparison may not be
appropriate.  Despite these caveats, there is a suggestion of
abundance evolution on average.

\section{MASS-METALLICITY RELATION} \label{sec:mzreln}

Galaxy mass is a somewhat more fundamental quantity than instantaneous
luminosity (i.e., it is more useful for predicting a galaxy's
properties).  The mass-metallicity relation is thus a better tool than
the luminosity-metallicity relation for quantifying the implications
of LIRG and ULIRG abundances for galaxy evolution.  However, mass has
its own set of limitations; for instance, mass is less easily measured
than luminosity.  In this section, we compare the masses and
abundances of LIRGs and ULIRGs to those of nearby star-forming
galaxies from the SDSS \citepalias{tremonti04a}.

The stellar masses for the ULIRGs are estimated from dynamical mass
measurements.  These are based on near-infrared measurements of
central stellar velocity dispersions and rotational velocities
\citep{dasyra06b}.  We do not have individual measurements for most
(2/3) of the ULIRG nuclei in our sample.  To plot individual nuclei on
the \mz\ relation, we instead use the high-precision average mass and
observed scatter of merger remnant nuclei from \citet{dasyra06b} to
assign masses to the ULIRGs in our sample.  The variance in mass is
computed from the mass equation and the observed scatter in random and
rotational velocities.  The mass of each galaxy is then drawn from a
Gaussian random distribution of the proper mean and variance, with the
limitation that the mass cannot deviate by more than 3$\sigma$ from
the mean.

Direct stellar mass measurements for LIRGs do not exist in the
literature.  However, the indirect method of modeling galaxy spectra
suggests an average stellar mass in star-forming LIRGs that is
one-half that of a ULIRG merger remnant ($\sim5\times10^{10}$~\msun;
\citealt{pkh05a}).  This is consistent with some LIRGs being
progenitor ULIRGs \citep{ishida04a}.  A caveat is that the
\citet{pkh05a} galaxies have log($L_\mathrm{FIR}$/\lsun) $=10.5-11.5$,
rather than the range log($L_\mathrm{IR}$/\lsun) $=11.0-12.0$ normally
assigned to LIRGs (note that $L_\mathrm{FIR} < \lir$ by $\sim$10\%; 
\citealt{sm96a}). Regardless, our comparison to \mz\ does not depend sensitively on our
choice of LIRG mass.  As with the ULIRGs, the mass of each galaxy is
drawn from a Gaussian random distribution of the proper mean and
variance.  We assume that the mean equals one-half the mean for ULIRGs
but that the variance in linear mass space is the same.

The SDSS galaxies to which we compare are primarily late-type
\citepalias{tremonti04a}, so their disks presumably have abundance
gradients.  The physical diameter spanned by the SDSS spectroscopic
fiber is relatively large (3\arcsec, corresponding to 4.6~kpc at
$\langle z \rangle = 0.08$).  Using trends of abundance dilution as a
function of aperture size gleaned from the NFGS data \citep{kjg05a},
we correct the SDSS abundances upward by 0.1~dex so that they better
approximate nuclear abundances.

\begin{figure}
  \plotone{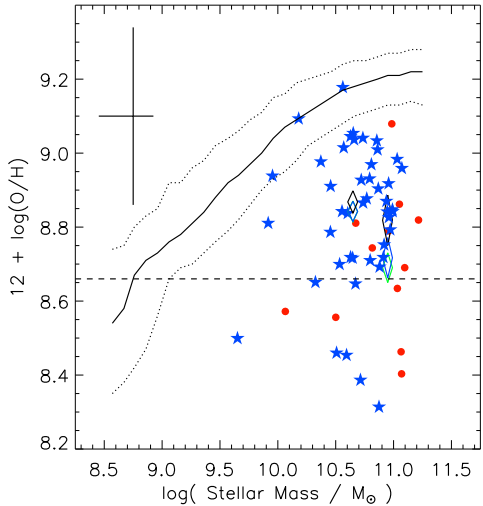}
  \caption{Comparison of the mass-metallicity relation from the SDSS
    \citep{tremonti04a} with LIRG and ULIRG abundances and stellar
    masses.  The average LIRG and ULIRG are significantly
    under-abundant, as they are when compared to the \lz\ relation.
    The dotted lines show 1$\sigma$ scatter on either side of the mean
    SDSS relation, which has been shifted upward by 0.1~dex to account
    for aperture effects.  Atop this relation are median LIRG and
    ULIRG abundances (colored diamonds).  The diamond colors represent
    different emission-line cuts (cut 1, green; cut 2, blue; cut 3,
    black), and the sizes represent the dispersion in points (the
    standard error in the median).  We also plot individual abundance
    measurements under the second emission-line cut, randomly
    distributed in mass according to the measured mean and standard
    deviation for LIRGs and ULIRGs \citep{pkh05a,dasyra06b}.}
  \label{fig_mz}
\end{figure}

The \mz\ relation is shown in Figure~\ref{fig_mz} with the LIRG and
ULIRG points over-plotted.  The results are comparable to those
obtained when we compare LIRGs and ULIRGs to the \lz\ relation.  LIRGs
and ULIRGs are significantly offset from the \mz\ relation, regardless
of the emission-line cut chosen.  We also observe a much larger
scatter in the abundances of LIRGs and ULIRGs than in the reference
sample at a similar mass.

\begin{figure}
  \plotone{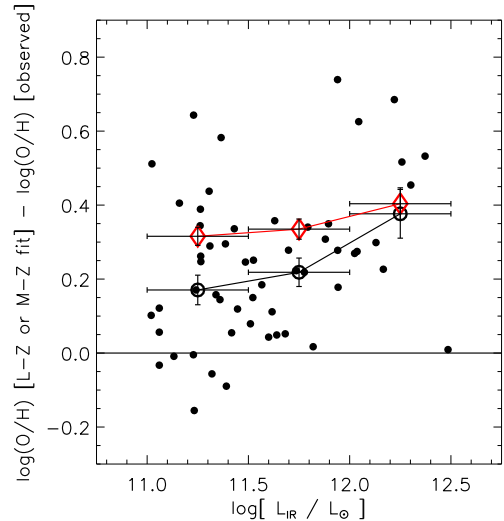}
  \caption{Difference between the observed abundances in LIRGs and
    ULIRGs and the \lz\ and \mz\ relations, as a function of infrared
    luminosity.  The (red) thick open diamonds are median deviations
    from the \mz\ relation.  There is excellent qualitative agreement
    between the two, though comparison to the \mz\ relation yields a
    slightly larger under-abundance of the LIRGs than does comparison
    to the \lz\ relation.  Refer to Figure \ref{fig_dzlzlir} for more
    details.}
  \label{fig_dzmzlir}
\end{figure}

In the case of \mz, the offsets from the mean relation are
unambiguously ones of abundance rather than mass, since the
progenitors of LIRGs and ULIRGs are massive spirals.  In
Figure~\ref{fig_dzmzlir}, we plot the abundance offsets of LIRGs and
ULIRGs from the \mz\ relation as a function of infrared luminosity
atop the same offsets computed using the \lz\ relation.  In each case
we use data that pass the second emission-line cut.  The median offset
of ULIRGs from the \mz\ relation, 0.4~dex, is the same as the offset
from the \lz\ relation (with our conservative flattening assumption),
but the offset from the \mz\ relation is higher for the LIRGs at
0.3~dex (vs.  0.2~dex from the \lz\ relation).  The correlation
between \lir\ and under-abundance may be present in the \mz\ offsets
as it is in the \lz\ offsets, but the observed trend is weaker and not
significant.  Nevertheless, given the systematic uncertainties in
abundance diagnostic calibrations, individual luminosity corrections,
individual masses, and the exact shape of the \lz\ relation, we
conclude that the \lz\ and \mz\ offsets are in good agreement with one
another.

\section{EFFECTIVE YIELD} \label{sec:effective-yield}

Using our measurements of the gas-phase abundances of oxygen in LIRGs
and ULIRGs, we are able to compute effective yields.  The true yield
$p$ is the fraction of the mass of a generation of stars that is
converted into a heavy element (in this case, oxygen) and returned to
the ISM.  More precisely, for a given stellar generation, $p$ refers
to the total mass of a heavy element produced by massive, short-lived
stars normalized by the mass locked up in long-lived stars and stellar
remnants.  The related quantity of effective yield is defined as
$p_{eff} \equiv Z~/~ln(\mu_g^{-1})$, where $\mu_g\equiv
M_{gas}/[M_{gas}+M_{stars}]$ is the gas mass fraction and $Z\equiv
M_{heavy~element}/M_{gas}$.  The effective yield provides information on
the chemical history of the galaxy through comparison with detailed
evolutionary models.  In the case of a `closed-box' model with
instantaneous recycling, the effective yield equals the true yield ($p
= p_{eff}$).

The effective yield is more sensitive to the chemical history of
galaxies than the \mz\ relation alone, since it also incorporates
information about the present gas content of the galaxy.  Star
formation increases a galaxy's effective yield until it asymptotically
reaches the true yield, by consuming gas and producing metals.
Conversely, gas flows in and out of the galaxy reduce the effective
yield \citep[e.g.,][]{edmunds90a,ke99a,dalcanton07a}.  An exception is
the case of inflows of gas with non-zero abundance.  Elementary models
of radial inflow in galaxy disks suggest that they can increase the
effective yield above the true yield by small amounts (factors of 2 or
less; \citealt{eg95a}).

A typical ULIRG gas mass fraction can be determined from global
\ion{H}{1} and H$_2$ observations.  Some uncertainty exists due to the
fact that the conversion of CO molecular luminosities to H$_2$ gas
masses differs between ULIRGs and normal galaxies and is not known
exactly \citep[cf.,][]{sss91a,ds98a}.  The uncertainty is compounded
by the scarcity of \ion{H}{1} measurements of ULIRGs.  We use the best
available data on ULIRG gas masses \citep{ms88a,ms89a,sss91a,ds98a},
and the average stellar mass of single nucleus ULIRGs from
\citet{dasyra06a}.  We then consider ranges of possible values for the
\ion{H}{1} mass fraction and CO-to-H$_2$ conversion factor, arriving
at an average gas mass fraction in ULIRGs of
$\langle\mu_g\rangle=0.1\pm0.05$.  (The conservative uncertainty
estimate is dominated by our uncertainty in $M_{HI}/M_{H_2}$.)  This
estimate is roughly consistent with the gas mass fraction estimated
from gas dynamical measurements ($M_{gas}/M_{dyn}\sim0.16$;
\citealt{ds98a}).  As expected, it is also much lower than the gas
mass fractions of high-redshift, submillimeter-selected ULIRGs
($\langle\mu_g\rangle=0.3-0.5$; \citealt{greve05a,tacconi06a}).

In the case of LIRGs, $M_{HI}/M_{H_2}$ is better constrained by
measurements \citep{ms89a,sss91a}.  The value of
$M_{H_2}/L^\prime_{CO}$ is also lower than in normal galaxies, though
its value relative to that in ULIRGs is uncertain.  An additional
uncertainty arises due to the absence of accurate stellar mass
measurements in LIRGs.  We resort to assuming that the stellar mass of
a LIRG is one-half that of a ULIRG (as in \S\ref{sec:mzreln}).  Again,
using the best measurements available and bracketing different
possibilities, we estimate $\langle\mu_g\rangle=0.2\pm0.1$.  (In this
case, the conservative uncertainty estimate is dominated by our
uncertainties in the stellar mass and $M_{H_2}/L^\prime_{CO}$.)

\begin{figure}
  \plotone{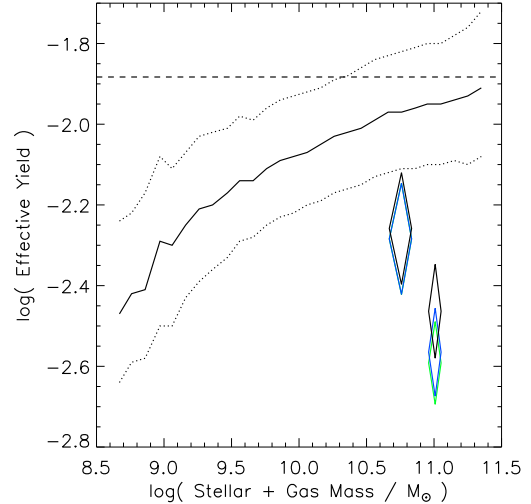}
  \caption{LIRG and ULIRG average effective yields compared to the
    mass-effective yield relation from the SDSS \citep{tremonti04a}.
    The average LIRG and ULIRG have effective yields that are lower
    than the average SDSS galaxy of the same masses by factors of 2
    and 3, respectively, as a consequence of abundance dilution.  The
    yields are computed using the median abundances for LIRGs and
    ULIRGs and gas mass fraction estimates of $\mu_g=0.1\pm0.05$ and
    $0.2\pm0.1$, respectively.  See Figure~\ref{fig_mz} for an
    explanation of most of the lines and symbols.  The horizontal
    dashed line is the SDSS estimate of the true yield.  The SDSS
    yields have been shifted upward by 0.1~dex to account for aperture
    effects.}
  \label{fig_myeff}
\end{figure}

Figure~\ref{fig_myeff} displays the median effective yields of oxygen
in LIRGs and ULIRGs compared to the \citetalias{tremonti04a}
mass-effective yield relation.  LIRGs and ULIRGs have significantly
lower effective yields, by factors of 2 and 3.5, respectively.  This
result mirrors the under-abundances described in previous sections,
since (a) $p_{eff}$ is linearly proportional to oxygen mass fraction
and (b) the LIRG and ULIRG gas mass fractions (0.2 and 0.1,
respectively) are similar to those of the average SDSS emission-line
galaxy of similar mass ($\mu_g=0.2$ for $M \sim 10^{11}~\msun$).  This
result is robust even if we have seriously underestimated $\mu_g$ in
ULIRGs: if $\mu_g$ in ULIRGs is raised to 0.2, $p_{eff}$ is still
significantly lower (by a factor of 2.5) in ULIRGs than in the SDSS
galaxies.

\section{DISCUSSION} \label{sec:discussion}

What is the cause of the \lz\ and \mz\ under-abundances
(Figures~$\ref{fig_lz1}-\ref{fig_lz3}$ and \ref{fig_mz})?  It is
accepted that the average progenitors of LIRGs and ULIRGs are gas-rich
spirals \citep{bh96a,mh96a,vks02a,iym04a,ishida04a,njb06a}.  Prior to
an interaction and the ensuing intense star formation episode, these
progenitor spirals presumably lie on the \lz\ and \mz\ relations.
Since star formation increases the abundance of a system, the only
processes that can reduce it are gas motions
\citep{edmunds90a,ke99a,dalcanton07a}.

Almost all LIRGs and ULIRGs host fast, massive, and powerful outflows
\citep{heckman00a,rvs02a,rvs05c,rvs05b,martin05a,martin06a}.  The hot
phase of these outflows is enriched in oxygen relative to iron
\citep[e.g.,][]{grimes05a}.  This enrichment arises because the hot
phase consists of supernovae ejecta from the current generation of
star formation.  Expulsion of these metal-rich ejecta therefore cannot
in principle reduce the abundance of the galaxy below its value prior
to the starburst.  Rather, outflow of hot gas prevents further
enrichment.

There exists also a neutral, much cooler gas phase in these outflows
with mass several times that in the hot phase
\citep{rvs05b}\footnote{\citet{rvs05b} write that the mean
  $M_{cold}/M_{hot}$ in LIRGs and ULIRGs is $\sim$0.5.  However, the
  normalization of the model that was compared to the observations is
  in error.  Correcting it implies that the actual entrainment value
  is $\sim$10 in LIRGs and $\sim$2 in ULIRGs, with an average of a few
  and a large scatter (from 0.25 to 20).}.  The average neutral gas
mass in LIRG and ULIRG winds is estimated to be 10$^9$~\msun\
\citep{rvs05b}, or $\sim$10\% of the average total gas mass.  The
effect on the galaxy's ISM is thus significant, and outflows could
remove the bulk of the galaxy's ISM over long timescales
(10$^8$-10$^9$ yr) if they act continuously.  The outflowing neutral
gas is swept up from the disk of the galaxy and has abundance equal to
that of the nuclear star-forming region.  Therefore, despite its large
mass, the neutral phase of LIRG and ULIRG outflows cannot reduce the
abundance of the nuclear regions, since it is not preferentially
removing heavy metals from the ambient gas.

Significant radial inflow of gas also occurs in gas-rich mergers
\citep[e.g.,][]{bh96a,mh96a,iym04a,njb06a}.  Transfer of angular
momentum from gas to stars through gravitational torques and
dissipation of kinetic energy in shocks causes much of the gas of the
progenitor systems to flow inward during the merger.  \citet{njb06a}
study the merger of two disk galaxies of equal mass, each with
$\mu_g=0.1$.  On the galaxies' first pass (the LIRG phase), $10-60\%$
of the gas in each disk flows inward to small radii (with the exact
value depending on geometry).  During the final coalescence (the ULIRG
phase), $50-85\%$ of the total gas moves to the nucleus.  The final
increase in the central gas mass is by a factor $3-4$ within a 2~kpc
radius \citep{iym04a}.

Radial abundance gradients exist in most disk galaxies, such that
abundances of oxygen and other elements decrease with increasing
distance from the galactic center \citep[e.g.,][]{zkh94a}.  Inflow of
gas consequently moves gas of low abundance into a nuclear region
whose initial abundance is higher.  As a result, the central regions
of LIRGs and ULIRGs would naturally exhibit lowered abundances due to
dilution.  The dilution of the nuclear abundance $Z$ is given by
$Z_{final}/Z_{initial} = (1+ab)/(1+a)$, where $a$ is the ratio of gas
mass arriving in the nucleus to that of gas already present and $b$ is
the ratio of abundances of the inflowing and nuclear gas.  For LIRGs
and ULIRGs we have shown that $Z_{final}/Z_{initial} \sim 0.5$, and
simulations predict $a=3-4$ \citep{iym04a}.  Inverting the above
equation and solving for $b$, we find $b=0.35$ (i.e., the inflowing
gas had 35\% of the initial abundance of the nuclear gas).  This is
consistent with the mean oxygen abundance gradient observed in spirals
(-0.6~dex per isophotal radius, or 25\% of the nuclear abundance at
the isophotal radius; \citealt{zkh94a}).

The effective yield of a galaxy can also change under a variety of
processes (\S\,\ref{sec:effective-yield}). However, in our case the
low effective yields we observe are primarily due to low abundances.
There is a secondary effect on $p_{eff}$ due to changes in $\mu_g$,
but the effect is small for these systems
(\S\ref{sec:effective-yield}).  Thus, the effective yields in LIRGs
and ULIRGs have been lowered primarily by the same processes that have
lowered the abundances.

The inflow scenario was proposed to explain the offset of local,
optically-selected interacting pairs from the $B$-band \lz\ relation
\citep{kewley06b}.  Our arguments above, based on reduced abundance
and effective yield, clearly make it the preferred explanation for
these infrared-selected mergers, as well.  As we discuss in
\S\ref{sec:comp-other-merg}, the offsets from the \lz\ relation in
LIRGs and ULIRGs are larger than those in optically-selected mergers.
Our interpretation implies that more gas has been funneled to the
centers of LIRGs and ULIRGs than in the optical mergers.  This
parallels the observation that ULIRGs are later-stage mergers and that
LIRGs and ULIRGs have higher star formation rates than the optical
mergers.  The amount of gas channeled to the center increases with
merger age \citep[e.g.,][]{njb06a}, and star formation is powered by
the gas driven to the merger center \citep[e.g.,][]{mh96a}.

The differences between the optical and infrared mergers could also
arise from their different chemical evolutionary histories following
inflow.  We have concluded that under-abundances in the
infrared-selected systems must be caused by gas inflow.  Following a
major inflow event, closed-box consumption of gas by star formation in
LIRGs and ULIRGs will tend to elevate the gas-phase abundances of
heavy elements, while strong outflows of metal-rich supernovae ejecta
will maintain low abundances.  Thus, the abundance dilution due to
inflow may have initially been stronger than observed, depending on
the subsequent effects of star formation and outflow.

The degree to which the initial under-abundance created by inflow was
smaller than the observed value can in principle be determined from
the properties of the ensuing star formation and outflow.  In
practice, the degeneracy among quantities such as the amount of gas
consumed by star formation, the amount of gas ejected by outflows, and
the enrichment of the outflowing gas means that such an exercise is
not well-constrained.  Simply from the good agreement between the
observed under-abundances and those expected from our basic
calculations of abundance dilution, we tentatively conclude that the
abundance immediately after inflow cannot be much lower than the
abundance observed now.  In Appendix~\ref{sec:appendix-a:-chemical},
we present example calculations that show subsequent abundance
evolution due to star formation and outflows need not undermine this
argument.  The uncertainties are too large to make quantitatively
useful statements about the impact of star formation and outflows.
However, these calculations do serve to illustrate that the scatter in
abundances we observe may be due to galaxy-to-galaxy differences in
how the inflow, star formation, and outflow have proceeded.

\section{SUMMARY AND OUTLOOK} \label{sec:summary-outlook}

We have shown that luminous and ultraluminous infrared galaxies do not
follow the standard luminosity-metallicity and mass-metallicity
relations for isolated galaxies with lower star formation rates.  As a
consequence, their effective yields are also smaller than those of
similar-mass galaxies.

We conclude that there is an oxygen under-abundance in the nuclei of
LIRGs and ULIRGs, compared to galaxies of the same luminosity and
mass.  We attribute this to radial inflow of gas into the galaxy
nuclei, as was concluded for studies of optically-selected mergers
\citep{kewley06b}.  This is consistent with the fact that many LIRGs
and all ULIRGs are in the early and late stages, respectively, of a
merger of two roughly equal-mass galaxies.  Other LIRGs are involved
in more minor interactions and/or mergers.  These interactions and
mergers cause gas from large radii, which is less oxygen-abundant than
the central regions, to fall to the center and dilute the nuclear
abundance.  This infalling gas also fuels star formation; LIRGs and
ULIRGs manifest the highest star formation rates in the local
universe.

Consistent with this interpretation is the observation that the LIRGs
and ULIRGs have larger offsets from the \lz\ relation ($0.2-0.4$~dex)
than optically-selected mergers of similar luminosity ($\la$0.1~dex;
\citealt{kewley06b}).  The LIRGs and ULIRGs are in more extreme
interaction states and/or have much higher star formation rates than
the optically-selected mergers.

A consequence of this result is that the \lz\ and \mz\ relations are
not universal.  At $z \ga 0.5$, where the star formation rate density
of the universe is dominated by LIRGs and ULIRGs
\citep[e.g.,][]{lefloch05a,caputi07a}, the \lz\ and \mz\ relations for
star-forming galaxies will be governed by mergers
(\citealt{smail04a,shi06a,bridge07a}, though see also
\citealt{mkl05a}).  Observations of high-redshift \lz\ and \mz\
relations that reveal a relation shifted to lower abundance and/or
higher luminosity and mass must account for this.  For instance, the
optically-selected galaxies at $z\sim2$ that were used to assess the
evolution of \mz\ are mostly LIRGs \citep{erb06a,reddy06a}.  A large
part of the offset attributed to evolution may be due to the fact that
the relations include more galaxies that are the by-product of
merging.  These high-$z$ mergers would have lower abundances not only
because they are younger but also because they have undergone gas
inflow.

Our low-redshift sample thus represents a baseline for comparison to
abundances of high-redshift LIRGs and ULIRGs.  The abundances of
high-redshift luminous infrared galaxies should be compared only with
the abundances of local galaxies with comparable star formation rate
and/or interaction strength, not with the abundances of field galaxies
of low star formation rate.  We compare our sample to a small sample
of LIRGs at $0.4 < z < 0.9$ \citep{liang04a} and find an increase in
abundance with decreasing redshift ($\sim$0.2~dex from $z\sim0.6$ to
$z=0.1$).  Three $z \sim 0.5$ ULIRGs from the current sample also have
lower abundances than the $z=0.1$ mean.  The situation for extremely
luminous, high-redshift SMGs \citep{swinbank04a,tecza04a,nesvadba07a}
is unclear, however; our upper limit on the mean abundance is
consistent with redshift evolution but more data are needed.  Many
more optical and near-infrared spectra of $z>0.5$ LIRGs and ULIRGs are
necessary to understand the abundance evolution of infrared-selected
mergers.

\acknowledgments

The authors are grateful to St\'{e}phane Charlot, Don Garnett, Lisa
Kewley, Stacy McGaugh, John Salzer, and Alice Shapley for helpful
discussion.  John Salzer kindly provided us with the KISS abundance
data for use in our figures.  Lisa Kewley also graciously provided an
IDL script which was the initial framework for computing the
abundances presented herein.  DSNR and SV were supported by the
following AST grants from the NSF: CAREER 9874973, ATI 0242860, and
EXC 0606932.

Some of the observations reported here were obtained at the W.~M. Keck
Observatory, which is operated as a scientific partnership among
Caltech, the University of California, and NASA; and the MMT
Observatory, which is a joint facility of the Smithsonian Institution
and the University of Arizona.  The Keck Observatory was made possible
by the generous financial support of the W.~M. Keck Foundation; we
thank those of Hawaiian ancestry on whose sacred mountain the
Observatory stands.  Our research has made use of IRAF, which is
distributed by the National Optical Astronomy Observatories; the
NASA/IPAC Extragalactic Database (NED), which is operated by
JPL/Caltech under contract with NASA; data products from the Two
Micron All Sky Survey, which is a joint project of the University of
Massachusetts and IPAC/Caltech and funded by NASA and NSF; and the
Sloan Digital Sky Survey, which is funded by the Alfred P. Sloan
Foundation, the Participating Institutions, the NSF, the U.S.
Department of Energy, NASA, the Japanese Monbukagakusho, the Max
Planck Society, and the Higher Education Funding Council for England.
The SDSS is managed by the Astrophysical Research Consortium for the
Participating Institutions.

\appendix
\section{Models of Post-Inflow Enrichment}
\label{sec:appendix-a:-chemical}

To illustrate interstellar chemical evolution in LIRGs and ULIRGs
subsequent to gas inflow, we present two example calculations, one
using a simple closed-box and the other a modified leaky-box model.
In these examples, we assume the true oxygen yield of $p = 0.0131$
from \citetalias{tremonti04a}, adjusted upward by 0.1~dex according to
our aperture corrections.  We refer to the oxygen abundance by mass
after inflow and before star formation and outflow as $Z_1$, and the
observed abundance as $Z_2$.  We use the average observed value of
$Z_2\sim0.006$ from this paper (in the \citetalias{tremonti04a}
calibration).  Similarly, the galaxy's gas masses before and after
star formation/outflow are $M_1$ and $M_2$.  For simplicity, we choose
a single value of $M_2/M_1=0.8$ as a reasonable guess (i.e., 20\% of
the galaxy's gas has been consumed in star formation and/or ejected by
outflows).  Exactly how much gas has been consumed by star formation
is unclear, but we do know that the gas-consumption timescale in
ULIRGs is short ($\la$100~Myr; \citealt{ds98a}).  Because our goal is
simply to illustrate that $Z_2/Z_1$ can be minimized, we leave it to5B
the reader to try other values of $M_2/M_1$.

In the closed-box model \citep{ta71a}, $Z_1$ is given by
\begin{equation}
  Z_1({\rm closed~box}) = Z_2 + p~{\rm ln}(M_2/M_1).
\end{equation}
Substituting the values listed above, we find $Z_1=0.003$, or half the
present-day value.  In the closed-box model with the values listed
above, consuming more than half of the gas present ($M_2/M_1\la0.5$)
implies that $Z_1<0$, which is unphysical.

In a leaky-box model \citep{hartwick76a}, outflows eject from the
galaxy some or all of the heavy elements produced by ongoing star
formation.  Two new quantities enter: $\eta \equiv
\frac{(dM/dt)_{OF}}{SFR}$ is the mass outflow rate normalized by the
star formation rate, and $\alpha \equiv Z_{OF}/Z_{ISM}$ is the
enrichment of the outflow relative to the ambient ISM.  For our
derivation, we begin with the treatment of \citet{matteucci01a}.  We
modify equation 5.33 of \citet{matteucci01a} by adding $\alpha$ to the
third term:
\begin{equation}
d(ZM_g)/dt = p~SFR-Z~SFR-\alpha~Z~(dM/dt)_{OF},
\end{equation}
where $Z$ is the ISM abundance, $M_g$ is the gas mass, SFR is the star
formation rate, and $(dM/dt)_{OF}$ is the mass outflow rate.  All
quantities but $p$, $\eta$, and $\alpha$ are assumed to be a function
of time.  Using the fact that $dM_g/dt=-(dM/dt)_{OF}-SFR$ (equation
5.32 of \citealt{matteucci01a}), we solve this equation through
substitution and integration.  The result is the equation for the
modified leaky-box case of an outflow with abundance greater than that
of the ambient ISM:
\begin{equation}
  Z_1({\rm leaky~box}, \alpha>1) = p/c-(M_2/M_1)^{-c/d}(p/c-Z_2),
\end{equation}
where $c\equiv\eta(\alpha-1)$ and $d\equiv\eta+1$.  Because star
formation and outflows cannot lower the galaxy's abundance (i.e., $Z_1
\leq Z_2$; see \S\ref{sec:discussion}), we have the further constraint
that $p/c-Z_2 > 0$, or $c \leq p/Z_2$ ($c\leq2.18$, for our values of
$p$ and $Z_2$).  In the case $\alpha=1$, the proper equation is the
standard leaky-box case of an outflow with abundance equal to that of
the ambient ISM:
\begin{equation}
  Z_1({\rm leaky~box}, \alpha=1) = Z_2 + (p/d)~{\rm ln}(M_2/M_1).
\end{equation}

\citet{rvs05b} estimate $\langle\eta\rangle\sim0.2-0.3$ (with a large
scatter) for the cool outflow phase of LIRGs and ULIRGs.  The cool
phase, which is mostly entrained ambient ISM, has $Z_{OF}\sim
Z_{ISM}$, or $\alpha=1$.  For star formation plus a cool-gas outflow,
$Z_1=0.0036-0.0039$, which is not far from the closed-box case.
Determining $\eta$ and $\alpha$ for the hot phase of LIRG and ULIRG
outflows is quite difficult, as they are poorly constrained
observationally.  By comparing the mass outflow rate in the cold phase
to predictions from starburst models, \citet{rvs05b} find $\eta$ could
be as high as $\sim$0.1.  Through model fits to the $\alpha$-element
enhancement, X-ray observations suggest $\alpha$ could reach values of
$\sim$10 \citep{grimes05a}.  These values yield $Z_1=0.0045$, 25\%
lower than the observed abundance.  If we combine the hot and cold
phases, we have $\eta\sim0.3-0.4$ and $\langle\alpha\rangle\sim4-4.3$,
which yields $Z_1=0.0046-0.0051$.  This is only $15-20$\% lower than
the average observed abundance in LIRGs and ULIRGs.

This exercise, though difficult because of observational
uncertainties, demonstrates two things: (a) the amount of post-inflow
enrichment required to reach the observed values of oxygen abundance
in LIRGs and ULIRGs need not be large, and (b) galaxy-to-galaxy
variations in chemical evolutionary histories following inflow events
can easily produce part of the observed scatter in abundances
(through, e.g., varying degrees of gas consumption, outflow
enrichment, and outflow efficiency).

\bibliography{apj-jour,dsr-refs}

\clearpage

\LongTables

\begin{deluxetable}{lllcrrcc}
  \tablecolumns{8}
  \tabletypesize{\footnotesize}
  \tablecaption{Sample\label{tab_samp}}
  \tablewidth{0pt}

  \tablehead{
    \colhead{{\it IRAS} FSC} & \colhead{Nuclear ID} &
    \colhead{Redshift} & \colhead{\lir} & \colhead{$K_s$} &
    \colhead{\%~Lum.} & \colhead{Sample} &
    \colhead{Ref} \\
    \colhead{(1)} & \colhead{(2)} & \colhead{(3)} & \colhead{(4)} &
    \colhead{(5)} & \colhead{(6)} & \colhead{(7)} & \colhead{(8)}
  }

  \startdata
\cutinhead{LIRGs}
           \objectname[2MASX~J00213417+3805347]{F00189+3748} &                        \objectname{2MASX~J00213417+3805347} &  0.036 & 11.34 &-25.06 &100.00 &  2jy &    3  \\
         \objectname[2MASXi~J0029256+303325]{F00267+3016:NW} &                         \objectname{2MASXi~J0029256+303325} &  0.050 & 11.73 &-25.39 & 76.80 &  2jy &    3  \\
        \objectname[2MASX~J00292498+3033339]{F00267+3016:SE} &                        \objectname{2MASX~J00292498+3033339} &  0.050 & 11.21 &-24.09 & 23.19 &  2jy &    3  \\
           \objectname[2MASX~J01200265+1421417]{F01173+1405} &                        \objectname{2MASX~J01200265+1421417} & 0.0312 & 11.63 &-24.38 &100.00 & rbgs &  423  \\
       \objectname[2MASX~J01385289$-$1027113]{F01364$-$1042} &                      \objectname{2MASX~J01385289$-$1027113} & 0.0484 & 11.76 &-23.88 &100.00 & rbgs &    4  \\
           \objectname[2MASX~J01511437+2234561]{F01484+2220} &                        \objectname{2MASX~J01511437+2234561} & 0.0323 & 11.64 &-25.50 &100.00 & rbgs &    5  \\
           \objectname[2MASX~J02274641+2635222]{F02248+2621} &                        \objectname{2MASX~J02274641+2635222} &  0.033 & 11.42 &-24.59 &100.00 &  2jy &    3  \\
                    \objectname[UGC~2369~NED01]{F02512+1446} &                \objectname{UGC~2369~NED01}\tablenotemark{a} & 0.0315 & 11.70 &-24.36 &100.00 & rbgs &   31  \\
                        \objectname[[CDF99\]~F1\_005]{F1\_5} &                                \objectname{[CDF99]~F1\_005} & 0.4786 & 11.85 & -1.00 &100.00 &  cdf &    1  \\
       \objectname[2MASX~J04340002$-$0834445]{F04315$-$0840} &                      \objectname{2MASX~J04340002$-$0834445} & 0.0160 & 11.60 &-24.74 &100.00 & rbgs &    6  \\
        \objectname[2MASX~J06573445+4624108]{F06538+4628:SW} &                        \objectname{2MASX~J06573445+4624108} &  0.021 & 11.31 &-24.03 &100.00 & rbgs &    3  \\
        \objectname[2MASX~J07091189+2036102]{IRAS07062+2041} &                        \objectname{2MASX~J07091189+2036102} &  0.017 & 11.13 &-24.17 &100.00 &  2jy &    3  \\
        \objectname[2MASX~J07091808+2038092]{IRAS07063+2043} &                        \objectname{2MASX~J07091808+2038092} &  0.017 & 11.26 &-24.80 &100.00 & rbgs &    3  \\
           \objectname[2MASX~J07285341+3349084]{F07256+3355} &                        \objectname{2MASX~J07285341+3349084} &  0.013 & 11.23 &-24.16 &100.00 & rbgs &    3  \\
              \objectname[IRAS~08572+3915NW]{F08572+3915:SE} &                              \objectname{IRAS~08572+3915NW} & 0.0583 & 11.23 &-22.17 & 13.57 &  1jy &    4  \\
           \objectname[2MASX~J09073082+1826057]{F09046+1838} &                        \objectname{2MASX~J09073082+1826057} &  0.029 & 11.31 &-24.78 &100.00 &  2jy &    3  \\
        \objectname[2MASX~J09155548+4419576]{F09126+4432:NE} &                        \objectname{2MASX~J09155548+4419576} & 0.0398 & 11.70 &-25.28 &100.00 & rbgs &    4  \\
         \objectname[2MASX~J09240034+3930426]{F09209+3943:E} &                        \objectname{2MASX~J09240034+3930426} & 0.0921 & 11.52 &-25.17 & 65.29 &  wgs &   24  \\
         \objectname[2MASX~J09235974+3930456]{F09209+3943:W} &                        \objectname{2MASX~J09235974+3930456} & 0.0922 & 11.25 &-24.49 & 34.70 &  wgs &    4  \\
            \objectname[FIRST~J092455.0+341535]{F09218+3428} &                         \objectname{FIRST~J092455.0+341535} &  0.068 & 11.72 & -1.00 &100.00 &  wgs &    2  \\
           \objectname[2MASX~J09355169+6121105]{F09320+6134} &                        \objectname{2MASX~J09355169+6121105} & 0.0393 & 11.96 &-25.53 &100.00 & rbgs &  423  \\
           \objectname[2MASX~J09363719+4828275]{F09333+4841} &                        \objectname{2MASX~J09363719+4828275} & 0.0259 & 11.37 &-24.20 &100.00 & rbgs &    3  \\
           \objectname[2MASX~J09364832+3119507]{F09338+3133} &                        \objectname{2MASX~J09364832+3119507} &  0.077 & 11.62 &-24.86 &100.00 &  wgs &    2  \\
           \objectname[2MASX~J09365478+2822200]{F09339+2835} &                        \objectname{2MASX~J09365478+2822200} &  0.119 & 11.78 &-24.96 &100.00 &  wgs &    2  \\
           \objectname[2MASX~J09425370+2816563]{F09399+2830} &                        \objectname{2MASX~J09425370+2816563} &  0.053 & 11.17 &-23.58 &100.00 &  wgs &    2  \\
                    \objectname[F10190+1322W]{F10190+1322:W} &                                   \objectname{F10190+1322W} & 0.0766 & 11.68 &-24.93 & 41.56 &  1jy &    1  \\
                    \objectname[F10190+1322E]{F10190+1322:E} &                                   \objectname{F10190+1322E} & 0.0759 & 11.82 &-25.27 & 58.44 &  1jy &   41  \\
           \objectname[2MASX~J10233258+5220308]{F10203+5235} &                        \objectname{2MASX~J10233258+5220308} & 0.0322 & 11.43 &-24.94 &100.00 &  2jy &   34  \\
                  \objectname[F10594+3818SW]{F10594+3818:SW} &                                  \objectname{F10594+3818SW} & 0.1578 & 11.94 &-24.68 & 38.03 &  1jy &    4  \\
           \objectname[2MASX~J11254505+1440359]{F11231+1456} &                        \objectname{2MASX~J11254505+1440359} &  0.034 & 11.57 &-24.81 &100.00 & rbgs &    3  \\
           \objectname[2MASX~J12140957+5431360]{F12116+5448} &                        \objectname{2MASX~J12140957+5431360} & 0.0081 & 11.06 &-23.09 &100.00 & rbgs &   64  \\
           \objectname[2MASX~J12273800+4009378]{F12251+4026} &                        \objectname{2MASX~J12273800+4009378} &  0.037 & 11.49 &-24.72 &100.00 &  2jy &    3  \\
           \objectname[2MASX~J13015026+0420005]{F12592+0436} &                        \objectname{2MASX~J13015026+0420005} & 0.0373 & 11.62 &-24.84 &100.00 & rbgs &    4  \\
           \objectname[2MASX~J13153506+6207287]{F13136+6223} &                        \objectname{2MASX~J13153506+6207287} &  0.031 & 11.79 &-24.18 &100.00 & rbgs & 4236  \\
           \objectname[2MASX~J13203537+3408218]{F13182+3424} &                        \objectname{2MASX~J13203537+3408218} &  0.022 & 11.67 &-23.98 &100.00 & rbgs &   56  \\
        \objectname[2MASX~J13395227+0050224]{F13373+0105:NW} &                        \objectname{2MASX~J13395227+0050224} &  0.022 & 11.39 &-24.82 &100.00 & rbgs &    6  \\
        \objectname[2MASX~J13395767+0049514]{F13373+0105:SE} &                        \objectname{2MASX~J13395767+0049514} &  0.023 & 11.15 &-25.04 &100.00 & rbgs &    6  \\
        \objectname[2MASX~J13481477+1525497]{F13458+1540:NE} &                        \objectname{2MASX~J13481477+1525497} &  0.058 & 11.43 &-24.96 & 49.07 &  2jy &    3  \\
           \objectname[CGCG~102$-$061~NED01]{F13458+1540:SW} &          \objectname{CGCG~102$-$061~NED01}\tablenotemark{a} &  0.058 & 11.45 &-25.00 & 50.92 &  2jy &    3  \\
           \objectname[2MASX~J14194323+4914121]{F14179+4927} &                        \objectname{2MASX~J14194323+4914121} &  0.026 & 11.32 &-23.97 &100.00 & rbgs &    5  \\
           \objectname[2MASX~J14301041+3112558]{F14280+3126} &                        \objectname{2MASX~J14301041+3112558} &  0.012 & 11.06 &-24.40 &100.00 & rbgs &    5  \\
           \objectname[2MASX~J14423488+6606043]{F14416+6618} &                        \objectname{2MASX~J14423488+6606043} &  0.038 & 11.27 &-23.38 &100.00 &  wgs &    2  \\
                    \objectname[F15043+5754S]{F15043+5754:S} &                                   \objectname{F15043+5754S} & 0.1505 & 11.94 &-25.44 & 49.54 &  1jy &    4  \\
        \objectname[2MASX~J15180612+4244445]{F15163+4255:NW} &                        \objectname{2MASX~J15180612+4244445} &  0.039 & 11.90 &-24.63 &100.00 & rbgs &    5  \\
           \objectname[2MASX~J15375657+3129582]{F15359+3139} &                        \objectname{2MASX~J15375657+3129582} & 0.0534 & 11.02 &-23.82 &100.00 &  wgs &   24  \\
           \objectname[2MASX~J15401502+3832111]{F15384+3841} &                        \objectname{2MASX~J15401502+3832111} & 0.0673 & 11.24 &-24.82 &100.00 &  wgs &   24  \\
            \objectname[FIRST~J154031.0+375814]{F15386+3807} &                         \objectname{FIRST~J154031.0+375814} & 0.1828 & 11.62 & -1.00 &100.00 &   ff &    1  \\
        \objectname[2MASX~J15410590+3204466]{F15391+3214:SE} &                        \objectname{2MASX~J15410590+3204466} & 0.0530 & 11.39 &-24.06 &100.00 &  wgs &   24  \\
                       \objectname[F15394+3532]{F15394+3532} &                                    \objectname{F15394+3532} & 0.1235 & 11.81 & -1.00 &100.00 &  wgs &   24  \\
           \objectname[2MASX~J15534897+3528042]{F15519+3537} &                        \objectname{2MASX~J15534897+3528042} & 0.0842 & 11.66 &-25.54 &100.00 &  wgs &   24  \\
            \objectname[FIRST~J155623.3+300443]{F15543+3013} &                         \objectname{FIRST~J155623.3+300443} &  0.121 & 11.71 & -1.00 &100.00 &  wgs &    2  \\
        \objectname[2MASX~J15560483+4149304]{F15543+4158:NW} &                        \objectname{2MASX~J15560483+4149304} &  0.134 & 11.74 &-25.55 & 74.96 &  wgs &    2  \\
         \objectname[FIRST~J155605.3+414922]{F15543+4158:SE} &                         \objectname{FIRST~J155605.3+414922} & 0.1342 & 11.26 &-24.53 & 25.03 &  wgs &   24  \\
           \objectname[2MASX~J15563641+4152501]{F15549+4201} &                        \objectname{2MASX~J15563641+4152501} & 0.0348 & 11.16 &-24.30 &100.00 &  wgs &   24  \\
           \objectname[2MASX~J16051287+2032326]{F16030+2040} &                        \objectname{2MASX~J16051287+2032326} &  0.015 & 11.02 &-23.38 &100.00 & rbgs &    5  \\
        \objectname[2MASX~J16114086+5227270]{F16104+5235:NE} &                        \objectname{2MASX~J16114086+5227270} & 0.0295 & 11.51 &-24.70 &100.00 & rbgs &   54  \\
                    \objectname[F16333+4630E]{F16333+4630:E} &                                   \objectname{F16333+4630E} & 0.1908 & 11.36 &-23.20 &  7.49 &  1jy &    1  \\
                    \objectname[F16474+3430N]{F16474+3430:N} &                                   \objectname{F16474+3430N} & 0.1126 & 11.43 &-23.82 & 15.28 &  1jy &    4  \\
           \objectname[2MASX~J16525886+0224035]{F16504+0228} &                        \objectname{2MASX~J16525886+0224035} & 0.0243 & 11.86 &-26.02 &100.00 & rbgs &    5  \\
        \objectname[2MASX~J16583138+5856102]{F16577+5900:SE} &                        \objectname{2MASX~J16583138+5856102} &  0.019 & 11.26 &-24.82 &100.00 & rbgs &    5  \\
           \objectname[2MASX~J18125527+6821484]{F18131+6820} &                        \objectname{2MASX~J18125527+6821484} &  0.020 & 11.23 &-24.83 &100.00 & rbgs &    6  \\
           \objectname[2MASX~J18431242+6039121]{F18425+6036} &                        \objectname{2MASX~J18431242+6039121} &  0.013 & 11.06 &-24.45 &100.00 & rbgs &    5  \\
         \objectname[2MASXi~J2057240+170735]{F20550+1655:SE} &                         \objectname{2MASXi~J2057240+170735} &  0.035 & 11.88 &-23.36 &100.00 & rbgs &    3  \\
    \objectname[2MASX~J22225728$-$0159290]{F22204$-$0214:NW} &                      \objectname{2MASX~J22225728$-$0159290} &  0.139 & 11.94 &-24.30 & 57.27 &  wgs &    2  \\
                                            F22204$-$0214:SE &                                    \nodata\tablenotemark{a} &  0.140 & 11.82 &-23.99 & 42.72 &  wgs &    2  \\
           \objectname[2MASX~J23511863+2034404]{F23488+2018} &                        \objectname{2MASX~J23511863+2034404} &  0.018 & 11.53 &-24.58 &100.00 & rbgs &    3  \\
\cutinhead{ULIRGs}
                   \objectname[F01298$-$0744]{F01298$-$0744} &                                  \objectname{F01298$-$0744} & 0.1361 & 12.41 &-24.38 &100.00 &  1jy &    1  \\
                       \objectname[F03250+1606]{F03250+1606} &                                    \objectname{F03250+1606} & 0.1290 & 12.19 &-25.93 &100.00 &  1jy &    1  \\
                   \objectname[F04313$-$1649]{F04313$-$1649} &                                  \objectname{F04313$-$1649} & 0.2672 & 12.72 &-25.14 &100.00 &  1jy &    1  \\
            \objectname[FIRST~J081645.7+310119]{F08136+3110} &                         \objectname{FIRST~J081645.7+310119} & 0.4070 & 12.46 &-26.30 &100.00 &   ff &    1  \\
                       \objectname[F08201+2801]{F08201+2801} &                                    \objectname{F08201+2801} & 0.1678 & 12.37 &-25.28 &100.00 &  1jy &    4  \\
            \objectname[FIRST~J082354.6+320212]{F08208+3211} &                         \objectname{FIRST~J082354.6+320212} & 0.3955 & 12.49 &-25.83 &100.00 &   ff &    1  \\
                       \objectname[F08474+1813]{F08474+1813} &                                    \objectname{F08474+1813} & 0.1454 & 12.28 &-24.18 &100.00 &  1jy &    1  \\
              \objectname[IRAS~08572+3915NW]{F08572+3915:NW} &                              \objectname{IRAS~08572+3915NW} &  0.058 & 12.04 &-24.09 & 86.43 &  1jy &   46  \\
                       \objectname[F08591+5248]{F08591+5248} &                                    \objectname{F08591+5248} & 0.1573 & 12.30 &-26.03 &100.00 &  1jy &    4  \\
                       \objectname[F09039+0503]{F09039+0503} &                                    \objectname{F09039+0503} & 0.1252 & 12.16 &-25.18 &100.00 &  1jy &   14  \\
                    \objectname[F09116+0334W]{F09116+0334:W} &                                   \objectname{F09116+0334W} & 0.1454 & 12.22 &-26.35 & 94.61 &  1jy &   14  \\
            \objectname[FIRST~J092734.7+324713]{F09245+3300} &                         \objectname{FIRST~J092734.7+324713} &  0.222 & 12.30 & -1.00 &100.00 &  wgs &    2  \\
                       \objectname[F09539+0857]{F09539+0857} &                                    \objectname{F09539+0857} & 0.1290 & 12.19 &-24.37 &100.00 &  1jy &    1  \\
                       \objectname[F10091+4704]{F10091+4704} &                                    \objectname{F10091+4704} & 0.2451 & 12.70 &-25.96 &100.00 &  1jy &    1  \\
            \objectname[FIRST~J101834.5+364951]{F10156+3705} &                         \objectname{FIRST~J101834.5+364951} & 0.4895 & 12.82 &-26.08 &100.00 &   ff &    1  \\
                       \objectname[F10494+4424]{F10494+4424} &                                    \objectname{F10494+4424} & 0.0919 & 12.23 &-25.05 &100.00 &  1jy &    4  \\
           \objectname[2MASX~J10591815+2432343]{F10565+2448} &                        \objectname{2MASX~J10591815+2432343} & 0.0430 & 12.04 &-25.03 &100.00 & rbgs &    3  \\
                       \objectname[F11028+3130]{F11028+3130} &                                    \objectname{F11028+3130} & 0.1986 & 12.49 &-24.65 &100.00 &  1jy &    4  \\
                       \objectname[F11387+4116]{F11387+4116} &                                    \objectname{F11387+4116} & 0.1489 & 12.26 &-25.50 &100.00 &  1jy &   14  \\
                       \objectname[F11506+1331]{F11506+1331} &                                    \objectname{F11506+1331} & 0.1274 & 12.39 &-25.39 &100.00 &  1jy &    4  \\
                  \objectname[F12112+0305NE]{F12112+0305:NE} &                                  \objectname{F12112+0305NE} &  0.073 & 12.00 &-24.21 & 45.64 &  1jy &  436  \\
                  \objectname[F12112+0305SW]{F12112+0305:SW} &                                  \objectname{F12112+0305SW} & 0.0730 & 12.08 &-24.40 & 54.36 &  1jy &   43  \\
                    \objectname[F13469+5833W]{F13469+5833:W} &                                   \objectname{F13469+5833W} & 0.1575 & 12.04 &-24.64 & 59.11 &  1jy &    4  \\
              \objectname[F14348$-$1447SW]{F14348$-$1447:SW} &                                \objectname{F14348$-$1447SW} &  0.083 & 12.15 &-25.18 & 62.18 &  1jy &    6  \\
                       \objectname[F15206+3342]{F15206+3342} &                                    \objectname{F15206+3342} & 0.1254 & 12.26 &-25.61 &100.00 &  1jy &   42  \\
           \objectname[2MASX~J15265942+3558372]{F15250+3608} &                        \objectname{2MASX~J15265942+3558372} & 0.0552 & 12.02 &-24.27 &100.00 & rbgs &   46  \\
            \objectname[FIRST~J154326.4+322859]{F15414+3238} &                         \objectname{FIRST~J154326.4+322859} & 0.2033 & 12.12 & -1.00 &100.00 &  wgs &   24  \\
           \objectname[2MASX~J16023279+3734532]{F16007+3743} &                        \objectname{2MASX~J16023279+3734532} &  0.185 & 12.09 &-26.18 &100.00 &  wgs &    2  \\
                    \objectname[F16333+4630W]{F16333+4630:W} &                                   \objectname{F16333+4630W} & 0.1908 & 12.45 &-25.85 & 92.51 &  1jy &    1  \\
                    \objectname[F17068+4027E]{F17068+4027:E} &                                   \objectname{F17068+4027E} & 0.1794 & 12.34 &-25.31 & 78.04 &  1jy &    1  \\
       \objectname[2MASX~J22404826$-$1321360]{F22381$-$1337} &                      \objectname{2MASX~J22404826$-$1321360} &  0.110 & 12.02 &-24.64 &100.00 &  wgs &    2  \\
                   \objectname[F22491$-$1808]{F22491$-$1808} &                                  \objectname{F22491$-$1808} &  0.076 & 12.17 &-24.92 &100.00 &  1jy &   26  \\
                    \objectname[F23234+0946W]{F23234+0946:W} &                                   \objectname{F23234+0946W} & 0.1279 & 12.07 &-25.41 & 78.72 &  1jy &    1  \\
           \objectname[2MASX~J23390127+3621087]{F23365+3604} &                        \objectname{2MASX~J23390127+3621087} &  0.065 & 12.13 &-25.22 &100.00 & rbgs &    2  \\
  \enddata

  \tiny{\tablerefs{1 = \citet{rvs02a,rvs05a}; 2 = \citet{kim95a}; 3 =
      \citet{wu98a}; 4 = SDSS; 5 = \citet{mk06a}; 6 = \citet{lk95a}}}

  \tiny{\tablecomments{Col.(1): Galaxy nucleus designation from the
      {\it IRAS} Faint Source Catalog.  Two sources are {\it IRAS}
      sources but not in the FSC, and a third is not an {\it IRAS}
      source (F1\_5; \citealt{cdf01a}).  The nucleus is specified if
      the IRAS flux is unresolved or spread among two or more nuclei.
      Col.(2): Designation, in NED nomenclature, which precisely
      specifies in sky coordinates the galaxy nucleus from which the
      IRAS flux and spectrum originate.  For 1~Jy ULIRGs, the nuclei
      are already known \citep{kvs02a,vks02a}.  For non-1~Jy ULIRGs
      and all LIRGs, we specify a unique galaxy nucleus designation.
      2MASS sources are used where possible.  Col.(3): Redshift.
      Sources are, in order of preference, the SDSS, \citet{rvs05a},
      \citet{kim95a}, \citet{wu98a}, or NED.  Col.(4):
      $8-1000$~\micron\ infrared luminosity.  The total galaxy
      luminosity has been multiplied by the factor in Column 5.
      Col.(5): $K_s$-band absolute magnitude, compiled and computed as
      described in \S\ref{sec:nir-photometry}.  Col.(6): Fraction of
      (unresolved) total infrared luminosity originating in this
      nucleus, based on $K$-band flux ratios for individual nuclei.
      Col.(7): Sample from which galaxy is taken: rbgs = Revised
      Bright Galaxy Sample \citep{sanders03a}, 1jy = the 1~Jy sample
      \citep{ks98a}, wgs = Warm Galaxy Sample \citep{kim95a}, 2jy =
      the 2~Jy sample \citep{strauss92a}, and ff = the FIRST/FSC
      sample \citep{stanford00a}.  Col.(8): Spectrum reference; the
      first reference is the data we use in our analysis.}}

  \tablenotetext{a}{A 2MASS designation exists for this galaxy, but it
    is attached to another nucleus.}

\end{deluxetable}

\clearpage

\begin{deluxetable}{lcrrllllllll}
\tablecolumns{10} 
\tabletypesize{\scriptsize}
\tablecaption{New Line Fluxes\label{tab_lines}}
\tablewidth{0pt}
\tablehead{
  \colhead{{\it IRAS} FSC} & \colhead{E(\bv)} &
  \colhead{$W_{eq}^{em}$(H$\alpha$)} &
  \colhead{$W_{eq}^{abs}$(H$\beta$)} & \colhead{\otw} &
  \colhead{H$\beta$} & \colhead{\ot} & \colhead{\oo} &
  \colhead{H$\alpha$} & \colhead{\nt} & \colhead{\st} & \colhead{\st}
  \\
  \colhead{} & \colhead{} & \colhead{} & \colhead{} &
  \colhead{3726$+$3729} & \colhead{4861} & \colhead{5007} &
  \colhead{6300} & \colhead{6563} & \colhead{6583} & \colhead{6716} &
  \colhead{6731} \\
  \colhead{(1)} & \colhead{(2)} & \colhead{(3)} & \colhead{(4)} &
  \colhead{(5)} & \colhead{(6)} & \colhead{(7)} & \colhead{(8)} &
  \colhead{(9)} & \colhead{(10)} & \colhead{(11)} & \colhead{(12)}
}

\startdata
\cutinhead{LIRGs}
     F01173+1405 &  0.93 & -202.00 &    3.29 &  9.83e-15 &  7.17e-15 &  6.67e-15 &  2.73e-15 &  6.02e-14 &  2.50e-14 &  1.05e-14 &  8.90e-15  \\
   F01364$-$1042 &  1.28 &  -18.30 &    9.84 &  1.83e-15 &  3.63e-16 &  3.94e-16 &  7.32e-16 &  3.19e-15 &  3.08e-15 &  1.27e-15 &  9.90e-16  \\
           F1\_5 &  0.91 & -119.48 &    6.13 &  1.49e-16 &  1.43e-16 &  5.56e-17 &   \nodata &  1.14e-15 &  5.13e-16 &  1.86e-16 &   \nodata  \\
  F08572+3915:SE &  0.44 &  -45.39 &    4.02 &  1.66e-15 &  8.70e-16 &  4.80e-16 &  1.20e-16 &  3.89e-15 &  1.15e-15 &  7.87e-16 &  5.89e-16  \\
  F09126+4432:NE &  1.45 &  -14.40 &    8.72 &  1.83e-15 &  6.29e-16 &  3.62e-16 &  3.80e-16 &  6.53e-15 &  6.14e-15 &  1.47e-15 &  1.07e-15  \\
   F09209+3943:W &  1.07 &  -11.03 &    5.10 &  8.19e-16 &  2.00e-16 &  2.09e-16 &   \nodata &  1.45e-15 &  9.87e-16 &  4.85e-16 &  3.52e-16  \\
     F09320+6134 &  1.63 &  -35.72 &    9.80 &  1.95e-15 &  1.02e-15 &  1.67e-15 &  1.07e-15 &  1.60e-14 &  1.92e-14 &  3.34e-15 &  2.53e-15  \\
   F10190+1322:W &  0.89 &  -36.10 &    6.14 &  9.21e-16 &  1.31e-15 &  2.16e-16 &  2.35e-16 &  9.31e-15 &  5.47e-15 &  1.08e-15 &  9.88e-16  \\
   F10190+1322:E &  1.88 &  -18.90 &   10.17 &  5.05e-16 &  1.57e-16 &  1.86e-16 &  3.43e-16 &  2.80e-15 &  2.41e-15 &  6.55e-16 &  4.80e-16  \\
  F10594+3818:SW &  0.51 &  -98.90 &    9.76 &  4.48e-15 &  3.10e-15 &  1.60e-15 &  6.97e-16 &  1.49e-14 &  7.74e-15 &  2.62e-15 &  2.05e-15  \\
     F12592+0436 &  1.34 &  -17.02 &    9.03 &  1.25e-15 &  4.46e-16 &  3.81e-16 &  5.46e-16 &  4.24e-15 &  3.63e-15 &   \nodata &   \nodata  \\
     F13136+6223 &  0.63 & -244.90 &    4.24 &  5.57e-14 &  3.00e-14 &  3.51e-14 &  5.64e-15 &  1.78e-13 &  6.93e-14 &  2.27e-14 &  2.10e-14  \\
   F15043+5754:S &  0.52 &  -80.16 &    5.82 &  2.36e-15 &  1.27e-15 &  7.97e-16 &  2.58e-16 &  6.28e-15 &  8.18e-16 &  1.18e-15 &  9.12e-16  \\
     F15386+3807 &  1.21 &  -75.33 &    4.84 &  1.56e-16 &  9.91e-17 &  8.59e-17 &  4.70e-17 &  1.11e-15 &  7.54e-16 &  2.00e-16 &  1.56e-16  \\
   F16333+4630:E &  0.49 &  -54.53 &    5.42 &  2.10e-16 &  9.50e-17 &  8.71e-17 &  2.50e-17 &  4.47e-16 &  1.73e-16 &  8.70e-17 &  6.31e-17  \\
   F16474+3430:N &  0.57 &  -66.99 &    7.80 &  2.12e-15 &  1.02e-15 &  6.19e-16 &  1.70e-16 &  5.20e-15 &  7.02e-16 &  9.81e-16 &  7.10e-16  \\
\cutinhead{ULIRGs}
   F01298$-$0744 &  0.92 &  -39.24 &    3.60 &  4.14e-16 &  2.26e-16 &  2.33e-16 &  1.80e-16 &  1.76e-15 &  9.46e-16 &   \nodata &   \nodata  \\
     F03250+1606 &  0.90 &  -23.10 &    6.95 &  7.92e-16 &  5.54e-16 &  5.65e-16 &  5.69e-16 &  3.61e-15 &  3.64e-15 &   \nodata &   \nodata  \\
   F04313$-$1649 &  1.66 &  -38.68 &    8.67 &  3.84e-16 &  2.23e-16 &  1.85e-16 &  1.22e-16 &  3.75e-15 &  9.30e-16 &  3.54e-16 &  2.71e-16  \\
     F08136+3110 &  0.75 & -118.41 &    6.72 &  5.90e-16 &  1.88e-16 &  2.10e-16 &  1.35e-16 &  1.23e-15 &  5.92e-16 &  3.40e-16 &  2.00e-16  \\
     F08201+2801 &  0.81 &  -80.65 &    7.71 &  1.88e-15 &  8.30e-16 &  6.51e-16 &  3.07e-16 &  5.71e-15 &  2.49e-15 &  1.06e-15 &  8.60e-16  \\
     F08208+3211 &  0.79 &  -95.11 &    2.69 &  2.67e-16 &  1.25e-16 &  1.41e-16 &  4.80e-17 &  8.83e-16 &  4.74e-16 &   \nodata &  1.42e-16  \\
     F08474+1813 &  1.56 &  -11.77 &    6.75 &  1.57e-16 &  3.30e-17 &  6.39e-17 &  5.15e-17 &  3.94e-16 &  5.44e-16 &  1.30e-16 &  1.09e-16  \\
  F08572+3915:NW &  0.69 &  -59.57 &    7.00 &  2.09e-15 &  1.25e-15 &  1.12e-15 &  4.60e-16 &  7.29e-15 &  2.89e-15 &  1.85e-15 &  1.35e-15  \\
     F08591+5248 &  0.82 &  -42.95 &    9.49 &  1.96e-15 &  9.42e-16 &  4.55e-16 &  4.32e-16 &  5.91e-15 &  4.40e-15 &  1.22e-15 &  9.98e-16  \\
     F09039+0503 &  0.54 &  -85.37 &    8.88 &  2.75e-15 &  2.07e-15 &  1.15e-15 &  1.03e-15 &  1.03e-14 &  7.52e-15 &  2.68e-15 &  2.56e-15  \\
   F09116+0334:W &  1.06 &  -25.11 &   11.69 &  2.30e-15 &  1.00e-15 &  5.46e-16 &  6.82e-16 &  7.15e-15 &  7.50e-15 &   \nodata &  1.35e-15  \\
     F09539+0857 &  0.97 &  -33.12 &    9.30 &  1.82e-15 &  3.19e-16 &  6.69e-16 &  4.88e-16 &  2.30e-15 &  3.65e-15 &  9.83e-16 &   \nodata  \\
     F10091+4704 &  1.09 &  -25.28 &    5.86 &  2.87e-16 &  6.95e-17 &  8.72e-17 &  1.06e-16 &  6.00e-16 &  6.24e-16 &  2.68e-16 &  1.94e-16  \\
     F10156+3705 &  0.21 & -229.17 &    8.46 &  7.44e-16 &  2.56e-16 &  3.75e-16 &   \nodata &  9.08e-16 &  3.50e-16 &   \nodata &  1.30e-16  \\
     F10494+4424 &  1.43 &  -50.70 &    7.00 &  8.37e-16 &  3.06e-16 &  4.72e-16 &  4.62e-16 &  4.21e-15 &  1.07e-15 &  1.07e-15 &  8.92e-16  \\
     F11028+3130 & -0.01 &   -8.44 &    9.09 &  4.01e-16 &  2.82e-16 &  8.91e-17 &   \nodata &  4.17e-16 &  3.77e-16 &  1.90e-16 &  1.05e-16  \\
     F11387+4116 &  1.02 &  -27.00 &    4.34 &  1.13e-15 &  4.95e-16 &  3.10e-16 &  4.00e-16 &  4.13e-15 &  4.10e-15 &  1.16e-15 &  1.00e-15  \\
     F11506+1331 &  1.13 &  -78.60 &    2.99 &  1.77e-15 &  6.16e-16 &  8.80e-16 &  5.76e-16 &  6.41e-15 &  2.74e-15 &  1.43e-15 &  1.13e-15  \\
  F12112+0305:NE &  0.89 &  -76.70 &    8.86 &  4.63e-15 &  1.48e-15 &  2.61e-15 &  1.07e-15 &  1.09e-14 &  4.62e-15 &  2.74e-15 &  1.99e-15  \\
  F12112+0305:SW &  1.33 &  -38.30 &    9.17 &  1.15e-15 &  3.55e-16 &  3.82e-16 &  4.93e-16 &  4.00e-15 &  2.60e-15 &  1.09e-15 &  8.07e-16  \\
   F13469+5833:W &  0.98 &  -26.02 &    6.44 &  6.73e-16 &  3.01e-16 &  2.17e-16 &  1.25e-16 &  2.25e-15 &  1.86e-15 &  4.80e-16 &  3.44e-16  \\
     F15206+3342 &  0.58 & -401.00 &    0.00 &  3.01e-14 &  1.93e-14 &  5.41e-14 &  3.38e-15 &  1.09e-13 &  2.04e-14 &  9.54e-15 &  7.87e-15  \\
     F15250+3608 &  0.75 &  -47.65 &    5.44 &  7.03e-15 &  3.11e-15 &  3.87e-15 &  1.49e-15 &  1.95e-14 &  9.34e-15 &  4.13e-15 &  3.32e-15  \\
   F16333+4630:W &  0.82 & -113.20 &    6.31 &  1.13e-15 &  5.52e-16 &  8.07e-16 &  3.58e-16 &  3.92e-15 &  2.46e-15 &  8.72e-16 &  7.13e-16  \\
   F17068+4027:E &  0.35 & -187.26 &   11.34 &  1.25e-15 &  8.40e-16 &  1.22e-15 &  2.21e-16 &  3.44e-15 &  1.07e-15 &  6.70e-16 &  5.40e-16  \\
   F23234+0946:W &  0.65 & -167.52 &    0.00 &  4.00e-15 &  1.45e-15 &  1.68e-15 &  1.02e-15 &  8.90e-15 &  9.90e-15 &  2.66e-15 &   \nodata  \\
\enddata

\tablecomments{Col.(1): Galaxy designation from the {\it IRAS} Faint
  Source Catalog.  Col.(2): Optical extinction, measured from the
  Balmer decrement.  Col.(3): H$\alpha$ emission line equivalent
  width, in \AA, uncorrected for underlying absorption.  Col.(4):
  H$\beta$ stellar absorption line equivalent width.  Col.(5-12):
  Observed line fluxes, in erg~s$^{-1}$~cm$^{-2}$.  Fluxes are
  uncorrected for extinction or underlying stellar absorption, except
  for H$\beta$ (which has had stellar absorption contamination
  removed).  Line flux measurement errors range are 5\%\ for strong
  lines, and up to $20-30$\%\ in a few individual cases for weak lines
  (e.g., \oo, \ot, and/or \st).}

\end{deluxetable}

\clearpage

\begin{deluxetable}{lrrcccccc}
\tablecolumns{9}
\tabletypesize{\footnotesize}
\tablecaption{Abundances\label{tab_abund}}
\tablewidth{0pt}

\tablehead{
  \colhead{IRAS FSC} & \colhead{log(\rtt)} & \colhead{log(\ott)} &
  \multicolumn{6}{c}{12$-$log(O/H)} \\
  \colhead{} & \colhead{} & \colhead{} & \colhead{1} & \colhead{2} &
  \colhead{3} & \colhead{4} & \colhead{5} & \colhead{6} \\
  \colhead{(1)} & \colhead{(2)} & \colhead{(3)} & \colhead{(4)} &
  \colhead{(5)} & \colhead{(6)} & \colhead{(7)} & \colhead{(8)} &
  \colhead{(9)}
}

\startdata
\cutinhead{LIRGs}
     F00189+3748 &  0.49 & -1.72 &    8.09 &    8.82 &    8.93 &    9.09 &    8.95 &    9.14  \\
  F00267+3016:NW &  1.36 & -1.60 & \nodata & \nodata & \nodata & \nodata & \nodata & \nodata  \\
  F00267+3016:SE &  0.91 & -1.01 &    7.70 &    8.33 &    8.44 &    8.68 &    8.84 &    9.10  \\
     F01173+1405 &  0.71 & -0.92 &    8.07 &    8.63 &    8.71 &    8.75 &    8.88 &    8.92  \\
   F01364$-$1042 &  1.37 & -1.84 & \nodata & \nodata & \nodata & \nodata & \nodata & \nodata  \\
     F01484+2220 &  0.33 & -1.10 &    8.38 &    8.94 &    9.04 &    8.98 &    9.04 & \nodata  \\
     F02248+2621 &  0.34 & -0.86 &    8.45 &    8.94 &    9.03 &    8.91 &    9.13 &    9.40  \\
     F02512+1446 &  0.65 & -1.14 &    8.07 &    8.69 &    8.79 &    8.85 &    8.92 &    9.07  \\
           F1\_5 &  0.54 & -1.25 &    8.16 &    8.79 &    8.89 &    8.92 &    8.95 &    9.06  \\
   F04315$-$0840 &  0.32 & -0.15 &    8.69 &    8.96 &    9.05 &    8.78 &    9.17 &    9.31  \\
  F06538+4628:SW &  0.71 & -1.24 &    7.95 &    8.61 &    8.72 &    8.86 &    8.88 & \nodata  \\
  IRAS07062+2041 &  0.33 & -0.78 &    8.49 &    8.95 &    9.04 &    8.90 &    9.06 &    8.97  \\
  IRAS07063+2043 &  0.72 & -1.41 &    7.88 &    8.58 &    8.70 &    8.89 &    8.94 & \nodata  \\
     F07256+3355 &  0.94 & -2.06 &    7.32 &    8.13 &    8.39 &    8.99 &    8.75 &    9.27  \\
  F08572+3915:SE &  0.59 & -1.05 &    8.17 &    8.75 &    8.84 &    8.85 &    8.84 &    8.76  \\
     F09046+1838 &  0.76 & -1.43 &    7.81 &    8.52 &    8.65 &    8.87 &    8.84 &    9.12  \\
  F09126+4432:NE &  1.21 & -2.02 & \nodata & \nodata & \nodata & \nodata & \nodata & \nodata  \\
   F09209+3943:E &  0.48 & -1.27 &    8.21 &    8.84 &    8.94 &    8.96 &    8.99 &    9.09  \\
   F09209+3943:W &  1.19 & -1.59 & \nodata & \nodata & \nodata & \nodata & \nodata & \nodata  \\
     F09218+3428 &  0.58 & -0.90 &    8.24 &    8.77 &    8.85 &    8.81 &    8.95 &    9.01  \\
     F09320+6134 &  1.16 & -1.29 & \nodata & \nodata & \nodata & \nodata & \nodata & \nodata  \\
     F09333+4841 &  0.90 & -1.24 &    7.62 &    8.32 &    8.45 &    8.76 &    8.84 &    9.13  \\
     F09338+3133 &  0.43 & -1.59 &    8.18 &    8.87 &    8.98 &    9.09 &    8.99 &    9.19  \\
     F09339+2835 &  0.56 & -1.09 &    8.19 &    8.78 &    8.87 &    8.87 &    8.96 &    9.05  \\
     F09399+2830 &  0.97 & -0.94 &    7.60 &    8.23 &    8.34 &    8.64 &    8.77 &    8.97  \\
   F10190+1322:W &  0.34 & -1.51 &    8.27 &    8.94 &    9.04 &    9.10 &    9.06 &    9.36  \\
   F10190+1322:E &  1.47 & -1.97 & \nodata & \nodata & \nodata & \nodata & \nodata & \nodata  \\
     F10203+5235 &  0.71 & -0.50 &    8.26 &    8.66 &    8.72 &    8.66 &    9.01 &    8.99  \\
  F10594+3818:SW &  0.51 & -0.98 &    8.28 &    8.83 &    8.91 &    8.87 &    9.00 &    9.09  \\
     F11231+1456 &  0.52 & -1.77 &    8.05 &    8.79 &    8.90 &    9.09 &    8.99 &    9.32  \\
     F12116+5448 &  0.55 & -0.36 &    8.48 &    8.82 &    8.88 &    8.71 &    9.04 &    9.15  \\
     F12251+4026 &  0.59 & -1.29 &    8.09 &    8.74 &    8.84 &    8.92 &    8.97 & \nodata  \\
     F12592+0436 &  1.15 & -1.69 & \nodata & \nodata & \nodata & \nodata & \nodata & \nodata  \\
     F13136+6223 &  0.73 & -0.74 &    8.13 &    8.62 &    8.69 &    8.70 &    8.87 &    9.03  \\
     F13182+3424 &  1.16 & -1.52 & \nodata & \nodata & \nodata & \nodata & \nodata & \nodata  \\
  F13373+0105:NW &  0.02 &  0.17 &    8.84 &    9.06 &    9.18 &    8.79 &    9.23 &    9.45  \\
  F13373+0105:SE &  0.58 & -0.85 &    8.26 &    8.77 &    8.85 &    8.81 &    8.67 & \nodata  \\
  F13458+1540:NE &  0.68 & -1.23 &    8.00 &    8.65 &    8.75 &    8.86 &    8.85 &    8.96  \\
  F13458+1540:SW &  0.44 & -0.68 &    8.45 &    8.89 &    8.97 &    8.83 &    9.04 &    9.04  \\
     F14179+4927 &  0.31 & -1.27 &    8.35 &    8.96 &    9.05 &    9.03 &    9.13 & \nodata  \\
     F14280+3126 &  0.37 & -1.49 &    8.24 &    8.92 &    9.02 &    9.08 &    8.99 &    9.29  \\
     F14416+6618 &  0.77 & -0.30 &    8.27 &    8.61 &    8.65 &    8.59 &    8.87 &    8.85  \\
   F15043+5754:S &  0.62 & -1.02 &    8.15 &    8.72 &    8.81 &    8.83 &    8.80 &    8.36  \\
  F15163+4255:NW &  0.69 & -0.87 &    8.12 &    8.66 &    8.74 &    8.75 &    8.93 & \nodata  \\
     F15359+3139 &  0.90 & -1.60 &    7.51 &    8.28 &    8.46 &    8.86 &    8.83 &    9.28  \\
     F15384+3841 &  0.51 & -1.09 &    8.24 &    8.83 &    8.92 &    8.90 &    8.94 &    9.37  \\
     F15386+3807 &  0.87 & -1.24 &    7.68 &    8.37 &    8.50 &    8.76 &    8.88 &    9.18  \\
  F15391+3214:SE &  0.71 & -1.78 &    7.81 &    8.57 &    8.72 &    9.01 &    8.88 &    9.28  \\
     F15394+3532 &  0.49 & -0.84 &    8.35 &    8.85 &    8.93 &    8.84 &    8.94 &    8.95  \\
     F15519+3537 &  0.85 & -1.30 &    7.70 &    8.41 &    8.54 &    8.79 &    8.84 &    9.27  \\
     F15543+3013 &  0.59 & -0.89 &    8.23 &    8.76 &    8.84 &    8.80 &    9.00 &    9.17  \\
  F15543+4158:NW &  0.57 & -0.69 &    8.34 &    8.79 &    8.87 &    8.77 &    8.95 &    8.72  \\
  F15543+4158:SE &  0.68 & -0.04 &    8.45 &    8.72 &    8.74 &    8.59 &    8.87 &    8.84  \\
     F15549+4201 &  0.76 & -0.69 &    8.10 &    8.58 &    8.65 &    8.68 &    8.91 &    9.04  \\
     F16030+2040 &  0.64 & -0.78 &    8.22 &    8.72 &    8.79 &    8.76 &    8.80 &    8.68  \\
  F16104+5235:NE &  0.38 & -0.82 &    8.44 &    8.92 &    9.01 &    8.89 &    9.06 & \nodata  \\
   F16333+4630:E &  0.71 & -0.88 &    8.10 &    8.64 &    8.72 &    8.75 &    8.86 &    8.90  \\
   F16474+3430:N &  0.68 & -1.15 & \nodata & \nodata & \nodata & \nodata & \nodata & \nodata  \\
     F16504+0228 &  1.05 & -1.28 & \nodata & \nodata & \nodata & \nodata & \nodata & \nodata  \\
  F16577+5900:SE &  0.61 & -1.29 &    8.07 &    8.73 &    8.83 &    8.91 &    8.95 &    9.07  \\
     F18131+6820 &  0.23 & -0.29 &    8.69 &    9.00 &    9.09 &    8.85 &    9.15 &    9.38  \\
     F18425+6036 &  0.45 & -0.97 &    8.33 &    8.87 &    8.96 &    8.89 &    9.10 &    9.23  \\
  F20550+1655:SE &  0.81 & -0.31 &    8.20 &    8.55 &    8.58 &    8.58 &    8.74 &    8.90  \\
F22204$-$0214:NW &  0.98 & -1.42 &    7.36 &    8.13 &    8.31 &    8.77 &    8.76 &    9.12  \\
F22204$-$0214:SE &  0.42 & -1.31 &    8.25 &    8.89 &    8.98 &    9.00 &    9.01 &    9.40  \\
     F23488+2018 &  0.60 & -1.26 &    8.09 &    8.74 &    8.84 &    8.90 &    8.97 &    9.23  \\
\cutinhead{ULIRGs}
   F01298$-$0744 &  0.82 & -1.02 &    7.87 &    8.48 &    8.58 &    8.73 &    8.86 & \nodata  \\
     F03250+1606 &  0.72 & -0.86 &    8.08 &    8.62 &    8.70 &    8.73 &    9.02 & \nodata  \\
   F04313$-$1649 &  1.11 & -1.65 & \nodata & \nodata & \nodata & \nodata & \nodata & \nodata  \\
     F08136+3110 &  0.95 & -1.16 &    7.55 &    8.24 &    8.38 &    8.71 &    8.76 &    8.88  \\
     F08201+2801 &  0.83 & -1.22 &    7.76 &    8.44 &    8.56 &    8.78 &    8.79 &    8.97  \\
     F08208+3211 &  0.83 & -0.96 &    7.87 &    8.47 &    8.56 &    8.71 &    8.87 &    8.99  \\
     F08474+1813 &  1.49 & -1.68 & \nodata & \nodata & \nodata & \nodata & \nodata & \nodata  \\
  F08572+3915:NW &  0.68 & -0.87 &    8.13 &    8.66 &    8.74 &    8.76 &    8.87 &    8.75  \\
     F08591+5248 &  0.78 & -1.46 &    7.78 &    8.50 &    8.63 &    8.88 &    8.91 &    9.22  \\
     F09039+0503 &  0.51 & -0.91 &    8.31 &    8.83 &    8.92 &    8.85 &    9.07 &    9.01  \\
   F09116+0334:W &  0.93 & -1.63 &    7.43 &    8.21 &    8.40 &    8.86 &    8.86 &    9.31  \\
     F09245+3300 &  0.62 & -0.86 &    8.21 &    8.73 &    8.81 &    8.77 &    8.99 &    9.13  \\
     F09539+0857 &  1.30 & -1.30 & \nodata & \nodata & \nodata & \nodata & \nodata & \nodata  \\
     F10091+4704 &  1.21 & -1.51 & \nodata & \nodata & \nodata & \nodata & \nodata & \nodata  \\
     F10156+3705 &  0.75 & -0.56 &    8.18 &    8.61 &    8.66 &    8.66 &    8.88 &    8.88  \\
     F10494+4424 &  1.21 & -1.40 & \nodata & \nodata & \nodata & \nodata & \nodata & \nodata  \\
     F10565+2448 &  0.91 & -1.24 &    7.60 &    8.31 &    8.44 &    8.75 &    8.86 &    9.22  \\
     F11028+3130 &  0.26 & -0.87 &    8.50 &    8.98 &    9.08 &    8.97 &    9.11 &    8.99  \\
     F11387+4116 &  0.92 & -1.51 &    7.50 &    8.26 &    8.43 &    8.83 &    8.91 &    9.18  \\
     F11506+1331 &  1.09 & -1.24 & \nodata & \nodata & \nodata & \nodata & \nodata & \nodata  \\
  F12112+0305:NE &  1.04 & -0.99 & \nodata & \nodata & \nodata & \nodata & \nodata & \nodata  \\
  F12112+0305:SW &  1.21 & -1.63 & \nodata & \nodata & \nodata & \nodata & \nodata & \nodata  \\
   F13469+5833:W &  0.90 & -1.39 &    7.58 &    8.31 &    8.46 &    8.80 &    8.87 &    9.26  \\
F14348$-$1447:SW &  0.57 &  0.01 &    8.57 &    8.82 &    8.86 &    8.64 &    9.12 &    8.95  \\
     F15206+3342 &  0.82 & -0.09 &    8.28 &    8.56 &    8.57 &    8.44 &    8.76 &    8.80  \\
     F15250+3608 &  0.84 & -0.90 &    7.88 &    8.45 &    8.54 &    8.69 &    8.83 &    8.92  \\
     F15414+3238 &  0.73 & -1.21 &    7.93 &    8.58 &    8.69 &    8.83 &    8.92 & \nodata  \\
     F16007+3743 &  1.06 & -1.46 & \nodata & \nodata & \nodata & \nodata & \nodata & \nodata  \\
   F16333+4630:W &  0.85 & -0.80 &    7.91 &    8.45 &    8.53 &    8.66 &    8.90 &    9.01  \\
   F17068+4027:E &  0.61 & -0.27 &    8.45 &    8.77 &    8.82 &    8.66 &    8.94 &    8.70  \\
   F22381$-$1337 &  0.61 &  0.01 &    8.53 &    8.78 &    8.82 &    8.52 &    8.95 &    8.82  \\
   F22491$-$1808 &  0.57 & -0.87 &    8.26 &    8.78 &    8.86 &    8.81 &    8.96 &    9.08  \\
   F23234+0946:W &  0.87 & -0.99 &    7.79 &    8.41 &    8.51 &    8.71 &    9.01 &    9.21  \\
     F23365+3604 &  0.64 & -1.14 &    8.08 &    8.70 &    8.79 &    8.85 &    9.00 &    9.03  \\
\enddata

\tablecomments{Col.(1): Galaxy designation from the {\it IRAS} Faint
  Source Catalog.  Col.(2): Logarithm of
  $\rtt\equiv\{f(\otwl)+f(\otll)\}/f(H\beta)$.  Col.(2): Logarithm of
  $\ott\equiv f(\otll)/f(\otwl)$.  Col.(4-9): Oxygen abundance from
  six different diagnostic/calibration systems.  1 = \citealt{pt05a}; 2 =
  \citealt{mcgaugh91a,dmd04a}; 3 = \citealt{tremonti04a}; 4 and 6 =
  \citealt{cl01a}, Cases F and A; and 5 = \citealt{kd02a}.  
  See \S\ref{sec:comp-diag} for a description of each system.}

\end{deluxetable}

\end{document}